\documentclass[journal=jctcce,manuscript=article]{achemso}

\usepackage{graphicx}
\usepackage{subfigure}
\usepackage{float}
\usepackage{dcolumn}

\usepackage{physics}
\usepackage[version=3]{mhchem}
\usepackage{algorithm}
\usepackage{algpseudocode}
\usepackage{amsmath}
\usepackage{amssymb}
\usepackage{bbm} 
\usepackage[super]{nth}

\usepackage{hyperref}
\hypersetup{
    colorlinks,
    linkcolor={blue},
    citecolor={blue},
    urlcolor={black}
}
\usepackage{color}

\newcommand{\ZHC}[1]{{#1}}

\newcommand{\eri}[2]{{\left( #1 \middle| #2 \right)}}

\newcommand*{\ii}{{\mathrm{i}}}
\newcommand*{\ee}{{\mathrm{e}}}
\newcommand*{\diff}{{\mathrm{d}} }

\newcommand{\pluseq}{\mathrel{+}=}
\newcommand{\timeseq}{\mathrel{*}=}

\newcommand*{\veck}{{\mathbf{k}}}
\newcommand*{\vecq}{{\mathbf{q}}}
\newcommand*{\vecG}{{\mathbf{G}}}
\newcommand*{\vecr}{{\mathbf{r}}}
\newcommand*{\vecR}{{\mathbf{R}}}
\newcommand*{\vecT}{{\mathbf{T}}}

\newcommand*{\veczero}{{\mathbf{0}}}
\newcommand*{\vecS}{{\mathbf{S}}}

\newcommand*{\imp}{{\mathrm{imp}}}

\newcommand*{\val}{{\mathrm{val}}}

\newcommand*{\LO}{{\rm LO}}
\newcommand*{\AO}{{\rm AO}}
\newcommand*{\EO}{{\rm EO}}
\newcommand*{\HL}{{\rm HL}}
\newcommand*{\LL}{{\rm LL}}

\newcommand*{\RPBE}{{\rm RPBE}}
\newcommand*{\RHF}{{\rm RHF}}

\newcommand*{\abinitio}{{\textit{ab initio}} }

\newcommand{\uroman}[1]{\uppercase\expandafter{\romannumeral#1}}
\newcommand{\lroman}[1]{\romannumeral #1}

\makeatletter
\DeclareRobustCommand\onlinecite{\@onlinecite}
\def\@onlinecite#1{\begingroup\let\@cite\NAT@citenum\citealp{#1}\endgroup}
\makeatother

\author{Zhi-Hao Cui}
\author{Tianyu Zhu}
\author{Garnet Kin-Lic Chan}
\email{gkc1000@gmail.com}
\affiliation{Division of Chemistry and Chemical Engineering, California Institute of Technology, Pasadena, CA 91125, USA}

\title{Efficient Implementation of Ab Initio Quantum Embedding in Periodic Systems: Density Matrix Embedding Theory}

\begin{tocentry}
\includegraphics[width=8.06cm,clip]{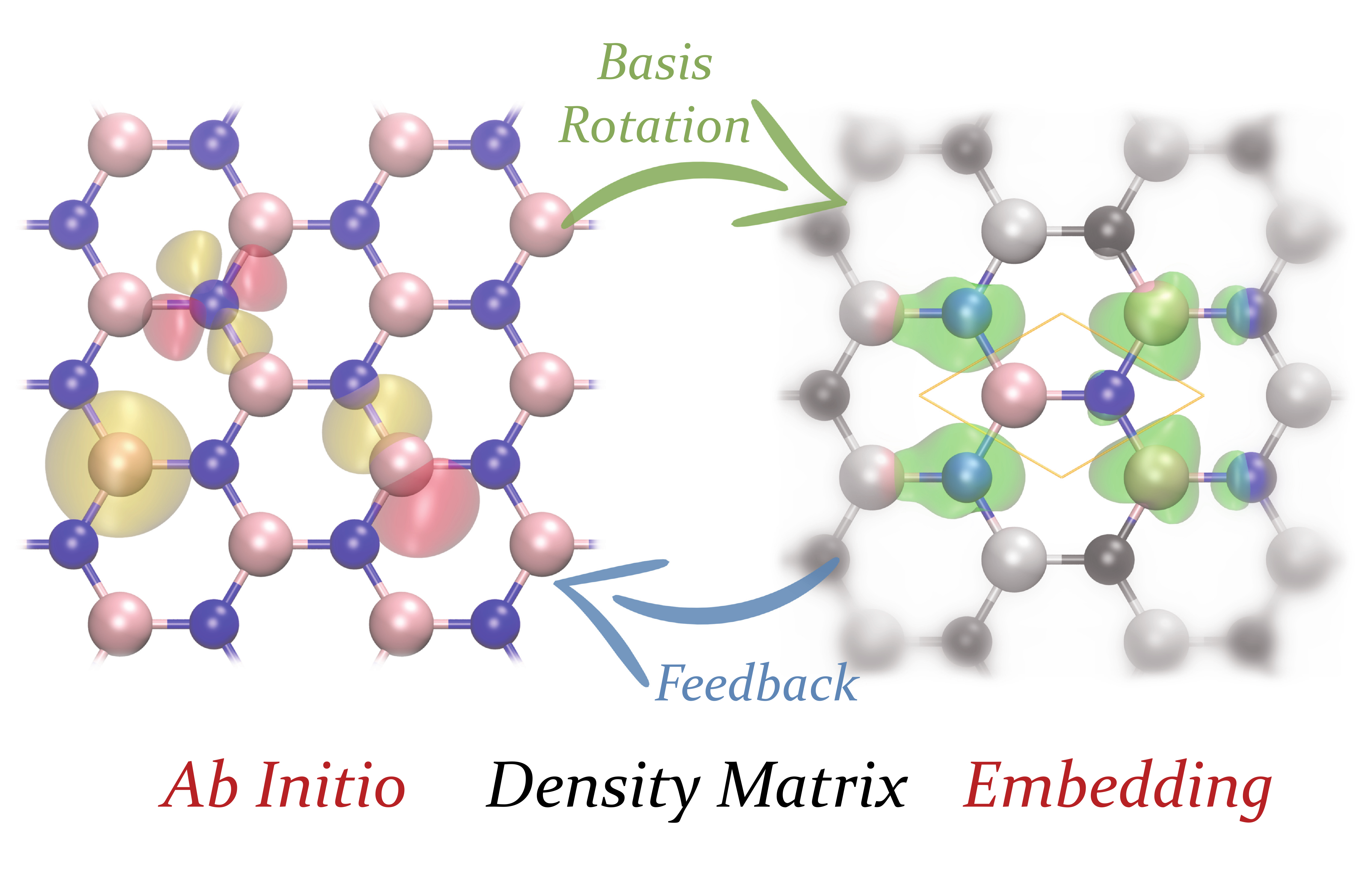}
\end{tocentry}

\begin{document}

\begin{abstract}
  We describe an efficient quantum embedding framework for realistic \abinitio density matrix embedding (DMET) calculations
  in solids.
  We discuss in detail the choice of orbitals and mapping to a lattice, treatment of the virtual space and bath truncation, and
  the lattice-to-embedded integral transformation. We apply DMET in this \abinitio framework  to a hexagonal boron nitride monolayer, crystalline silicon,
  and nickel monoxide in the antiferromagnetic phase, using large embedded clusters with up to 300 embedding orbitals.
  We demonstrate our formulation of \abinitio DMET in the computation of ground-state properties such as the total energy, equation of state, magnetic moment and correlation functions.
\end{abstract}

\section{Introduction}\label{sec:intro}

The \abinitio description of strongly correlated electrons in solids is a major challenge,
limiting the quantitative understanding of interacting electronic phases, such as the Mott~\cite{Imada98} and high-temperature superconducting
phases~\cite{Dagotto94, Sachdev03, Lee06RMPhightc}. The heart of the difficulty lies in the need to use
computational methods that can treat correlated electrons, which usually means a steep computational scaling with system size,
as well as treat the thermodynamic limit (TDL), in order to observe distinct phases.


A formal route to extend high-level correlated electron methods to infinite systems is provided by \emph{quantum embedding}~\cite{Zgid11, Sun16QET}.
While there are today a wide variety of techniques termed embedding~\cite{Sun16QET}, 
we will be concerned with the type of quantum embeddings in condensed phases that historically started with the treatment of defects in solids via the Anderson impurity model, where the
interacting impurity site is surrounded by a set of bath orbitals that approximately represent the environment \cite{Anderson61}. 
This impurity idea can be generalized to translationally invariant systems, where the lattice is subdivided into multiple clusters (also termed impurities or fragments) where each is embedded in a self-consistent environment generated by the other impurities.
In the embedding treatment, only the solution of the embedded cluster (i.e. the cluster along with its quantum bath)
is treated by the high-level correlated method (the impurity solver), while interactions between clusters
are treated at a lower level of theory, typically within a single-particle framework such as mean-field. 


Dynamical mean-field theory (DMFT) was the first  quantum embedding algorithm for periodic
systems based on the above self-consistent quantum impurity idea~\cite{Georges92, Georges96}, and has since been
extended in many different directions and settings~\cite{Georges96, Kotliar06RMP, Held2007, Maier05RMP, Potthoff03, Senechal08, Kananenka15, Rusakov19, Biermann14JPCM}.
DMFT is formulated in terms of the one-particle Green's function,
and solving the embedded impurity problem yields a local self-energy that is then used
in the single-particle Green's function description of the periodic lattice. 
More recently, density matrix embedding theory (DMET) \cite{Knizia12} has been proposed as a computationally simpler
quantum embedding algorithm, also for a self-consistent quantum impurity, but adopting the one-particle reduced density matrix as the
fundamental variable, in conjunction with a static mean-field description of the periodic lattice~\cite{Knizia12,
Bulik14, Chen14dmethoneycomb, Zheng16, Zheng17}. Because DMET only requires to compute frequency-independent observables, 
it is less expensive than DMFT, and in practice, a wider variety of correlated electron methods can be applied to the impurity problem.
A further kind of quantum embedding, density functional (or wavefunction-in-density functional) embedding
\cite{Wesolowski93, Goodpaster10dftemb, Huang11dftemb, Libisch14, Jacob14, Chulhai2018, Lee19dftemb, Zhu2016, Zhu2019a} is also of much current interest.
However, this is not usually applied to strongly correlated phases, and thus we do not consider it further here.


In this work, we will focus our attention on the \abinitio implementation of DMET in periodic solids.
While DMET has been successfully applied to compute electronic phase diagrams across a range of strongly correlated
lattice models~\cite{Knizia12, Bulik14, Chen14dmethoneycomb, Fan15, Zheng16, Zheng17, Zheng17sci, Gunst17, Sandhoefer16,
Wu19pdmet},  the extension
of DMET to a practical \abinitio method for periodic systems remains incomplete. There have been several works on
cyclic H and Be ring structures\cite{Knizia13, Wouters16, Fulde17dmet, Pham18} and an early DMET implementation for solids that
treated minimal unit cells and small basis sets~\cite{Bulik14detsolid} (e.g. 2D boron nitride in the 6-31G basis and diamond in a STO-3G basis~\cite{Bulik14detsolid}). However, such calculations are best considered
model \abinitio calculations in the sense that the basis sets and impurity sizes are too small for
quantitative or chemical accuracy.
What remains to be developed is a comprehensive computational framework in periodic DMET calculations that can use
both large and realistic basis sets, and treat non-trivial cluster sizes or complicated unit cells with many atoms.
Describing such a framework is the purpose of the current work.

To establish a practical implementation of \abinitio periodic DMET, it is worth outlining the similarities and differences between a
calculation on a lattice model and a realistic solid. On the one hand, both models and real solids are translationally invariant over cells, and thus
for an efficient computational algorithm, $\veck$-point symmetry should be utilized wherever possible. On the other hand, there are many
important differences, i.e. (\lroman{1}) in a realistic solid, one needs to define the impurity basis, and different
definitions can vary widely in terms of locality and other properties, (\lroman{2}) the number of atoms and basis functions per impurity cell
can be very large in a realistic system, and (\lroman{3}) realistic Hamiltonians contain complicated interactions between
all the basis functions, including potentially divergent long-range Coulomb terms. Thus, realizing \abinitio DMET  involves both specifying some, in principle,
arbitrary choices (such as the choice of impurity orbitals) as well as carrying out efficient implementations of many standard
quantum chemistry routines, such as integrals and their transformations. The latter is also part of the general infrastructure of
\abinitio periodic quantum chemistry. In this work we rely heavily on 
the periodic computational infrastructure established in the \textsc{PySCF} package~\cite{Sun18pyscf, McClain17, Sun17},
which in fact historically grew out of an effort to implement \abinitio DMET.


The remainder of the paper is organized as  follows. In Sec. \ref{sec:theory}, we first describe the detailed DMET embedding framework for periodic solids, including the definition of the impurity and lattice basis, the construction of local orbitals, bath truncation, efficient
integral transformation, and DMET and charge self-consistency. 
In Sec. \ref{sec:results}, we apply the method to some prototype crystals with realistic basis sets and
non-trivial cluster sizes with up to $\sim 300$ embedded cluster orbitals, including a 2D hexagonal boron nitride monolayer, 3D crystalline silicon, and the antiferromagnetic (AFM) \uroman{2} phase of NiO. We finish in Sec. \ref{sec:conclusion} with conclusions and remarks. 

\ZHC{Note added: In a recent submission, Pham et al. have also presented related work \cite{Pham19DMETsolid} that applies \abinitio DMET to
  periodic systems.}

\section{Theory}\label{sec:theory}

\subsection{DMET Implementation}\label{subsec:dmet}

In this section, we describe the detailed implementation of DMET for \abinitio calculations in solids, focusing
on aspects related to periodic systems that have not been reported in the previous DMET literature.
For a general description of the DMET algorithm (and a detailed description of its molecular implementation)
we refer readers to Ref. \onlinecite{Wouters16}. 

{\bf Lattice and impurity localized orbitals}. The infrastructure of
\abinitio mean-field theory uses crystal (Bloch) orbitals and $\veck$-point quantities, while quantum embedding is naturally formulated
in terms of local orbitals and real-space quantities. Thus, we first define a translation 
from the mean-field computational basis to one appropriate for embedding. 

To do so, we construct atom-centered orthogonal local orbitals (LO) $\qty{w_i (\vecr)}$ that define the lattice Hilbert space, which can be cleanly partitioned
into a product of impurity Hilbert spaces.
Here, we will assume that the mean-field computational basis is a set of crystal
atomic orbitals (AOs) $\qty{\phi^{\veck}_{\mu} (\vecr)}$ (which constitutes a non-orthogonal basis, with an AO index $\mu$
and a $\veck$-point index in the first Brillouin zone). 
It is convenient to first define an intermediate set of local crystal orbitals,
\begin{equation}\label{eq:C aolo}
  w^{\veck}_{i} (\vecr) 
  = \sum_{\mu} \phi^{\veck}_{\mu} (\vecr)C^{\veck, \AO, \LO}_{\mu i} ,
\end{equation}
where the notation $C^{\mathrm{X}, \mathrm{Y}}$ denotes the transformation from basis $\mathrm{X}$ to basis $\mathrm{Y}$.
The real-space LOs in any cell can then be obtained by a Wannier summation over the local crystal orbitals, for example,
the LOs at the lattice origin  ($\vecR = \veczero$) are given by
\begin{equation}\label{eq:w0}
w^{\vecR = \veczero}_{i} (\vecr) = \frac{1}{\sqrt{N_{\veck}}} \sum_{\veck} w^{\veck}_{i} (\vecr).
\end{equation}
Expressed in the LOs,  the \abinitio periodic system is isomorphic to a periodic lattice problem, with reciprocal lattice vectors $\veck$.
We choose a subset of $\qty{w_i(\vecr)}$ to define the impurity. It is natural to choose the
impurity to be spanned by LOs in a single unit cell or a supercell, and
for definiteness, we choose the cell or supercell at the lattice origin as the impurity.

{\bf{Choice of local orbitals.}} The next computational task is to specify the coefficients in Eq.~\ref{eq:C aolo} that define
the LOs in terms of the crystal AOs. 
There are two strategies to construct orthogonal local orbitals:
a \emph{top-down} strategy [transforming from  canonical mean-field molecular orbitals (MOs) to LOs] and a \emph{bottom-up} strategy (transforming from the AO computational basis to LOs).
The first strategy finds a unitary transformation of the MOs to
optimize a metric (such as $\expval{r^2} - \expval{\vecr}^2$) that measures the spatial locality of the LOs. Examples of such approaches
are the Boys\cite{Foster60}, Pipek-Mezey (PM)\cite{Pipek98} and Edmiston-Ruedenberg
(ER)\cite{Edmiston63}  methods in molecules, and the maximally localized Wannier function
(MLWF)\cite{Marzari97, Marzari12RMP} and Pipek-Mezey Wannier function (PMWF)\cite{Jonsson17} methods in solids.
The top-down scheme can yield more localized orbitals than bottom-up schemes. However,
due to the need to carry out an optimization, the disadvantages are also apparent: (\lroman{1})
the procedure can be numerically expensive and one can easily get stuck in a local minimum of
the cost function, particularly when constructing a large number of local
virtual orbitals; (\lroman{2}) with periodic boundary conditions, entangled bands \cite{Souza01, Damle18} often exist among the high-energy virtual MOs, and special techniques
are required; (\lroman{3}) a false minimum or discontinuity in $\veck$-space can lead to non-real orbitals
after the Wannier summation in Eq. \ref{eq:w0}, giving a Hamiltonian with complex coefficients in the LO basis,
which is incompatible with many impurity solver implementations. 


In the bottom-up strategy, one avoids optimization and relies only on linear algebra to construct the LOs.
Examples of LOs of this type are the L\"owdin and meta-L\"owdin orbitals \cite{Lowdin50, Sun14qmmm}, natural atomic orbitals (NAO) \cite{Reed85} and
intrinsic atomic orbitals (IAO) \cite{Knizia13IAO}. 
Bottom-up methods avoid the difficulties of the top-down strategy: (\lroman{1}) the construction is usually
cheap (i.e. suited to producing large numbers of local orbitals); (\lroman{2}) there is no initial guess dependence or
local minimum problem; (\lroman{3}) the LOs are guaranteed to be real as long as the phases of crystal AOs and other $\veck$-space orbitals in the formalism (e.g. the reference crystal AOs used to construct the IAOs) are smooth in $\veck$-space. Since we aim to carry out calculations beyond a minimal basis, and thus with many virtual
orbitals, we have chosen the bottom-up strategy to avoid difficulties in optimization and non-real Hamiltonian coefficients. In
particular, we have adapted the molecular IAO routine to crystal MOs with $\veck$-point sampling (see Appendix~\ref{app: kiao})
to generate the set of crystal IAOs.
The crystal IAOs are \emph{valence} orbitals that exactly span the occupied space of the mean-field calculation.
Note that the number of IAOs is the same as the size of the minimal basis only. To obtain a complete
set of LOs that span the same space as the original AO basis 
(thus making a square rotation matrix $C^{\veck, \AO, \LO}$ in Eq. \ref{eq:C aolo}) 
we need to further augment the IAOs with LOs that live purely in the virtual space.
Here we choose these additional orbitals to be the projected atomic orbitals (PAO) for non-valence orbitals
\cite{Saebo93}, orthogonalized with L{\"o}wdin orthogonalization, as originally proposed for local correlation calculations\cite{Saebo93}.
The IAOs + PAOs then together span the complete space of AOs and constitute a complete LO basis.
A related scheme has previously been used in the molecular DMET calculations\cite{Wouters16, Motta17}.

{\bf{DMET bath and truncation.}} The DMET embedded Hilbert space consists of the impurity LOs and a set of bath orbitals; these together
are the embedding orbitals (EOs). We define the bath orbitals in DMET by using the SVD of the
mean-field off-diagonal density matrix between the impurity and remaining lattice $\gamma^{\vecR \neq \veczero, \veczero}_{ij}$~\cite{Wouters16}, 
\begin{equation}\label{eq:bath R SVD}
\gamma^{\vecR \neq \veczero, \veczero}_{ij} = \sum_{\tilde{i}} B^{\vecR \neq \veczero}_{i\tilde{i}} \Lambda_{\tilde{i}\tilde{i}} V^{\veczero \dagger}_{\tilde{i}j} .
\end{equation}
where $B^{\vecR \neq \veczero}$ gives the coefficients of the bath orbitals and we use ``$\sim$'' above the orbital indices
to denote orbitals in the embedding space.
The overall projection from the LO basis to the EO basis then has the following form,
\begin{equation}\label{eq:C loemb R}
C^{\vecR, \LO, \EO} = \begin{bmatrix} 
\mathbbm{1}    & \veczero \\
\veczero & \mathbf{B}^{\vecR \neq \veczero}
\end{bmatrix} ,
\end{equation}
where the identity block means that the impurity LOs (i.e. the basis defined in Eq. \ref{eq:w0}) are left unchanged.
To transform from the computational crystal AO basis to the embedding orbitals, we multiply two transformations,
\begin{align}\label{eq:C loemb k}
  C^{\veck, \LO, \EO} = \sum_{\vecR} \ee^{-\ii \veck \cdot \vecR} C^{\vecR, \LO, \EO} \notag, \\
  C^{\veck, \AO, \EO} = C^{\veck, \AO, \LO} C^{\veck, \LO, \EO} .
\end{align}



Although the DMET bath is formally of the same size as the number
of impurity orbitals, the mean-field wavefunction only contains appreciable entanglement
between partially occupied LOs on the impurity and corresponding bath orbitals. Very low-lying core
and high-energy virtual impurity orbitals thus are not entangled with any bath orbitals.
In practice, this manifests as very small singular values $\Lambda_{\tilde{i}\tilde{i}}$ 
and the corresponding singular vectors (bath orbitals) can vary between different DMET iterations~\cite{Wouters16}
leading to difficulties in converging the DMET self-consistency procedure.
To eliminate this instability, we use the procedure previously recommended in molecular DMET calculations
\cite{Wouters16}.
We first partition the impurity orbitals into  core, valence and virtual orbitals, and only carry out the SVD
for the impurity valence columns of the off-diagonal density matrix to construct corresponding valence bath
orbitals~\cite{Wouters16}, i.e. the index $j$ in Eq. \ref{eq:bath R SVD} can be constrained to the valence orbitals only.
Note that when pseudopotentials are used in the calculation, there is no core subspace, and thus no core bath
orbitals appear. With this construction, the
number of embedding orbitals is reduced from $2 n_{\imp}$ to $n_{\imp} + n_{\val}$, where $n_{\val}$ is the number of valence
orbitals, which is smaller than the number of impurity orbitals $n_{\imp}$, and we recover smooth DMET convergence.


{\bf{Constructing the embedding Hamiltonian.}} Using the EOs defined above, we can construct the DMET embedding Hamiltonian.
The embedding Hamiltonian in the DMET  \emph{interacting bath} formalism\cite{Knizia13,Wouters16} takes the form,
\begin{equation}\label{eq:emb Ham}
\mathcal{H} = \sum_{\tilde{i}\tilde{j}} \tilde{F}_{\tilde{i}\tilde{j}} c^{\dagger}_{\tilde{i}}c_{\tilde{j}} - \mu \sum_{\tilde{i}\in \imp} c^{\dagger}_{\tilde{i}}c_{\tilde{i}} +\frac{1}{2} \sum_{\tilde{i}\tilde{j} \tilde{k}\tilde{l}} \eri{\tilde{i}\tilde{j}}{\tilde{k}\tilde{l}} c^{\dagger}_{\tilde{i}}c^{\dagger}_{\tilde{k}}c_{\tilde{l}}c_{\tilde{j}} . 
\end{equation}
Besides the normal one- and two-particle terms, a chemical potential $\mu$ is added to the impurity Hamiltonian so that the number of electrons
on the impurity is constrained to be precisely correct. An alternative choice is
the DMET \emph{non-interacting bath} formalism~\cite{Wouters16}. In this case, the two-particle interactions
are restricted to the impurity orbitals, and interactions on the bath are mimicked by adding the
 correlation potential to the bath. For further details, we refer to Ref.~\onlinecite{Wouters16}. 
 In this work, we primarily use the interacting bath formalism, and only briefly consider the non-interacting bath formalism for comparison.


To obtain the coefficients of the embedding Hamiltonian, we first transform the Fock matrix from the AOs to the EOs,
\begin{equation}\label{eq:fock transform}
F^{\veczero, \EO} = \frac{1}{N_{\veck}}\sum_{\veck} C^{\veck, \AO, \EO \dagger} F^{\veck, \AO}  C^{\veck, \AO, \EO} ,
\end{equation}
where $F^{\veck, \AO}$ is the Fock matrix in the periodic mean-field calculation. [Note that regardless
of the mean-field orbitals used (i.e. Hartree-Fock or DFT), the Fock matrix refers to the Hartree-Fock one-particle
Hamiltonian, \emph{not} the Kohn-Sham Hamiltonian]. To eliminate double counting, we subtract the contribution
of the embedding electron repulsion integrals (ERIs, see below for their construction) from
the transformed Fock matrix $F^{\veczero, \EO}$ in Eq. \ref{eq:fock transform},
\begin{equation}\label{eq:fock double counting}
\tilde{F}_{\tilde{i}\tilde{j}} = F^{\veczero, \EO}_{\tilde{i}\tilde{j}} - \qty[\sum_{\tilde{k}\tilde{l}} \eri{\tilde{i}\tilde{j}}{\tilde{k}\tilde{l}} \gamma_{\tilde{l}\tilde{k}} - \frac{1}{2}\eri{\tilde{i}\tilde{k}}{\tilde{l}\tilde{j}} \gamma_{\tilde{k}\tilde{l}}] ,
\end{equation}
where $\gamma$ is the density matrix rotated to the embedding basis. 

The construction and integral transformation of the two-particle ERIs of the embedding orbitals can be computationally expensive.
A significant reduction in cost is obtained by using density fitting \cite{Whitten73, Sun17}. Density fitting defines the 4-center ERIs
in terms of the 3-center ERIs. In the presence of  $\veck$ symmetry, this takes the form
\begin{equation}\label{eq:density fitting}
\eri{\mu \veck_{\mu} \nu \veck_{\nu}}{\kappa \veck_{\kappa} \lambda  \veck_{\lambda}} \approx \sum_{L} \eri{\mu \veck_\mu \nu \veck_\nu}{L} \eri{L}{\kappa \veck_\kappa \lambda \veck_\lambda} ,
\end{equation}
where $L$ is the auxiliary basis and only three $\veck$ indices are independent.
There are many choices of auxiliary basis and here we will mainly use Gaussian density fitting (GDF), where $L$ is a 
set of chargeless Gaussian crystal orbitals, with the divergent part of the Coulomb term
treated in Fourier space~\cite{Sun17}. [We discuss plane-wave density fitting (FFTDF) in Appendix \ref{app:fftdf eri}]. $L$ has an implicit $\veck$ dependence in Eq.~\ref{eq:density fitting}. This means the 3-center integral $\eri{L}{\mu\veck_\mu \nu\veck_\nu}$ is more precisely written as $\eri{L\veck_L}{\mu\veck_\mu \nu\veck_\nu}$, where $\veck_L=\veck_\mu-\veck_\nu + n\mathbf{b}$ due to momentum conservation ($n\mathbf{b}$ is integer multiple of reciprocal lattice vectors). 
We construct the embedding ERIs starting from the GDF 3-center integrals according to Algorithm \ref{alg:eri with gdf}.
\begin{algorithm}[hbt]
  \caption{Pseudocode for the embedding ERI transformation with GDF.} 
  \label{alg:eri with gdf}
  \begin{algorithmic}[1]
    \For{all $\veck_{L}$}
    \For{$\qty(\veck_{\mu}, \veck_{\nu})$ that conserves momentum} 
    \State Transform $\eri{L}{\mu\veck_\mu \nu\veck_\nu}$ to $\eri{L}{\tilde{i}\veck_\mu \tilde{j} \veck_\nu}$ by $C^{\veck, \AO, \EO}$
    \Comment{$\veck$-AO to $\veck$-EO}
    \State $\eri{L}{\veczero \tilde{i} \veczero \tilde{j}} \pluseq \frac{1}{N_{\veck}} \eri{L}{\tilde{i} \veck_\mu \tilde{j} \veck_\nu}$ \Comment{FT to the reference cell $\vecR = \veczero$}
    \EndFor
    \State $\eri{\tilde{i}\tilde{j}}{\tilde{k}\tilde{l}} \pluseq \frac{1}{N_{\veck}} \sum_{L} \eri{\veczero \tilde{i} \veczero \tilde{j}}{L} \eri{L}{\veczero \tilde{k} \veczero \tilde{l}}$
    \Comment{Contraction for the embedding ERI}
    \EndFor
  \end{algorithmic}
\end{algorithm}
\ZHC{In this algorithm, the final contraction step scales as $\mathcal{O}\qty(n_{\veck} n_{L} n^4_{\EO})$ while the transformation step ($\veck$-AO to $\veck$-EO) scales as  $\mathcal{O}\qty(n^{2}_{\veck} n_{L} n_{\AO} n^{2}_{\EO}) + \mathcal{O}\qty(n^{2}_{\veck} n_{L} n^2_{\AO} n_{\EO})$ where we use $n_{\AO} (n_{\EO})$ to denote the number of atomic (embedding) basis functions per cell. Note that $n_{\EO}$ is larger than $n_{\AO}$ and
thus the first term is the dominant term.} If the number of $\veck$-points is not too large, the contraction
is the rate determining step. It is noteworthy that the scaling with respect to $\veck$
is only linear (contraction) and quadratic (transformation).
    As an example, the embedding ERIs of a $3\times3\times1$ cluster of boron nitride (GTH-DZVP basis and a $6\times6\times1$ mean-field lattice 
    corresponding to transforming 936 crystal AOs to 306 embedding orbitals) can be constructed in about 200s using 28 cores.
    The largest objects during the calculation are the final set of ERIs $\eri{\tilde{i}\tilde{j}}{\tilde{k}\tilde{l}}$ and the AO density fitting integral $(L|\mu\veck_\mu \nu \veck_\nu)$.  The latter is stored on disk and loaded into memory blockwise to further reduce the required memory.
    Finally, we note that if the impurity solver supports density fitting without requiring explicit ERIs, the contraction step in Algorithm \ref{alg:eri with gdf} can be omitted. 

{\bf{DMET and charge self-consistency.}} A key component in the DMET description of phases and order parameters
is the imposition of self-consistency between the ``high-level'' (HL) embedded wavefunction and the ``low-level'' (LL)
mean-field description. We matched the correlated one-particle density matrix $\gamma$ from the impurity solver and
the mean-field one-particle density matrix by minimizing their 
Frobenius norm difference with respect to the correlation potential $u$,
\begin{equation}\label{eq:cost func}
\min_{u} \sum_{ij \in {\EO}} \qty[\gamma^{\LL}_{ij}(u) - \gamma^{\HL}_{ij}]^2 ,
\end{equation}
where the indices $i, j$ loop over all embedding orbitals \ZHC{and the high-level density matrix $\gamma^{\HL}_{ij}$ is kept fixed during the correlation potential fitting. } Other choices of cost function are also possible, e.g. only matching the impurity \cite{Knizia13, Wouters16} or diagonal part \cite{Bulik14} of the density matrix. However, we only consider full matching in this work. \ZHC{The correlation potential
  is a local term (i.e. independent of the impurity cell $\veck$).  In the current work, the correlation potential is chosen
  to be a spin-dependent potential where the number of independent elements per spin-component is $n_{\LO} (n_{\LO} + 1) / 2$.} With large basis sets, the number of parameters in $u$ can be very large. To reduce the degrees of freedom in the numerical optimization, we can add $u$ only to a subset of orbitals, e.g. the valence orbitals. With a small set of parameters, the optimization problem can be easily solved, e.g. by a conjugate gradient algorithm. \ZHC{It should be noted that the minimization of the cost function is not a convex problem, thus
  in principle there can be multiple local minima; for example
  in an AFM system, there may be multiple solutions corresponding to different spin polarization patterns. However, we have not observed multiple local minima in this work, since the BN and Si systems do not break spin symmetry, and in NiO, we always start with a particular AFM order in the
  initial guess of $u$.}

In an \abinitio DMET calculation, an additional layer of self-consistency appears associated
with the non-linear \abinitio lattice mean-field calculation [this is sometimes referred to as \emph{charge self-consistency} (CSC)
in DMFT calculations \cite{Savrasov01, Savrasov04, Pourovskii07, Park14dmftcsc}]. In our implementation, the AO-based Fock matrix $F^{\veck, \AO}$ is updated at the beginning of each DMET cycle, using the improved DMET mean-field density matrix from the previous iteration, which reflects the response of the mean-field density (matrix) to the DMET local correction. We always perform CSC in our calculations unless otherwise specified.

We finally note that the LOs, in principle, can be redefined based on the new mean-field MOs at each DMET iteration.
However, we do not consider such an update in the current work. Instead, we only determine the LOs at the beginning of
the calculation and keep the LOs fixed in the following DMET self-consistency loops. This choice introduces a small
dependence on the initial orbitals (e.g. using HF- or DFT-MOs to define the LOs). However, it is usually reasonable to assume that
the LOs do not change significantly during the embedding self-consistency. 

\begin{figure}[hbt]
\includegraphics[width=0.5\textwidth]{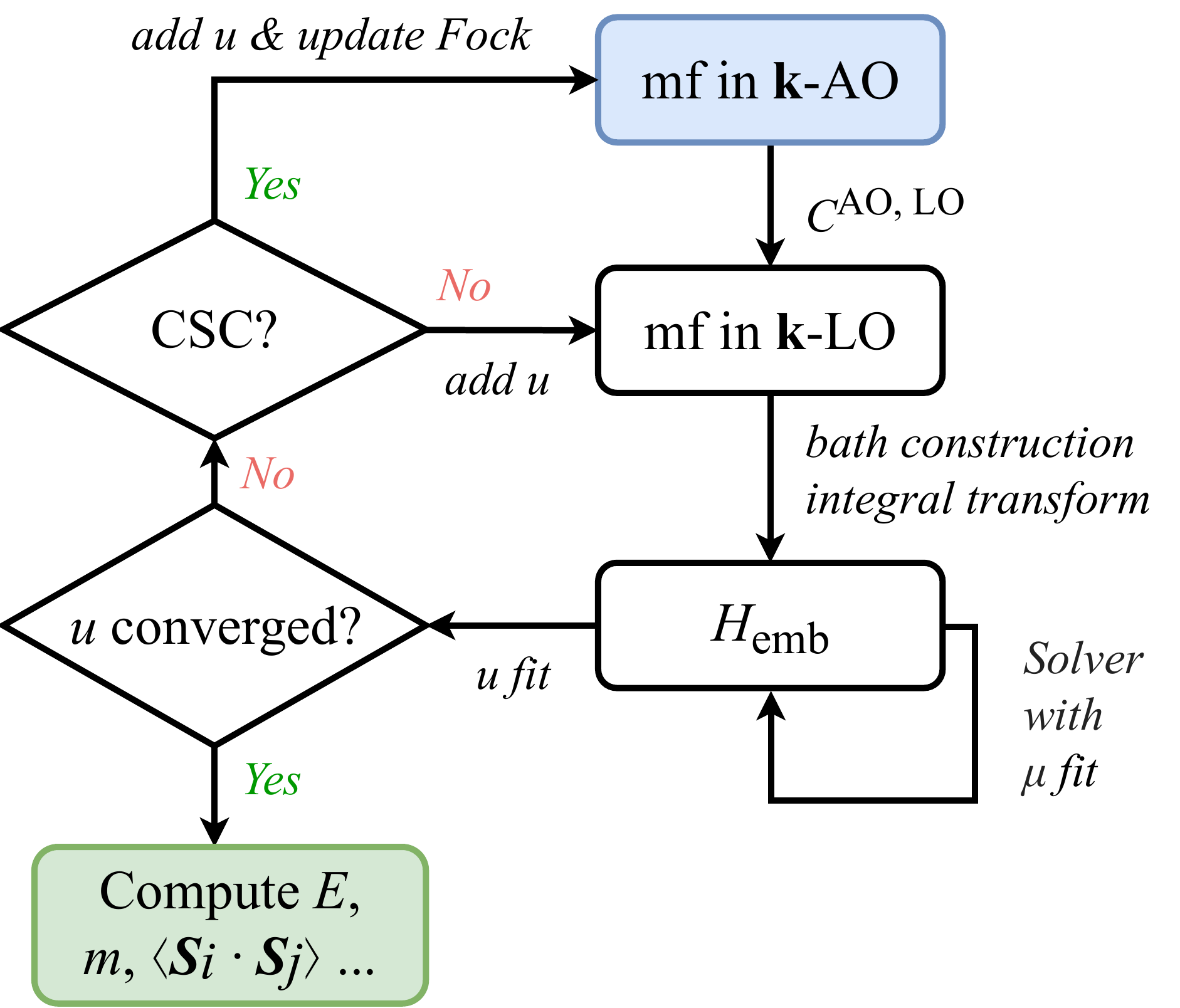}
\caption{\ZHC{The DMET self-consistency procedure, where  ``mf'' is used to denote the relevant mean-field physical quantities, e.g. the Fock matrix $F$, density matrix $\gamma$; $\mu$ and $u$ are used to denote the chemical potential and correlation potential respectively. ``CSC'' denotes charge self-consistency and is an optional step in the algorithm. The flowchart starts at the blue block and ends at the green block when self-consistency is reached. }}\label{fig:DMET flowchart}
\end{figure}

\ZHC{
We illustrate the periodic \abinitio DMET algorithm, with both DMET correlation potential and charge self-consistency, in Fig. \ref{fig:DMET flowchart}. 
}

\subsection{Computational Details}\label{sec:compdetails}

We consider three prototypical solids: a 2D hexagonal boron nitride monolayer (h-BN), crystalline silicon (Si) and nickel monoxide (NiO). The lattice parameters were taken from experiment: $a = 2.50 \text{\AA}$ for the BN monolayer\cite{Li11hBN} (with $20.0 \text{\AA}$ vacuum to eliminate fictitious interactions between mirrors); $a = 5.43053 \text{\AA}$ for Si \cite{Tobbens01Si}, and $a = 4.17 \text{\AA}$ for NiO \cite{Cheetham83NiO}. To target the AFM-\uroman{2} state, the minimal unit cell of NiO was chosen as the rhombohedral cell that contains two formula units of NiO. \ZHC{We used 28 Intel E5-2680@2.40GHz cores in all the calculations.} We summarize the computational parameters for DMET below.

{\bf{Mean-field calculations.}} All mean-field calculations were performed using the \textsc{PySCF} package \cite{Sun18pyscf} with Hartree-Fock or DFT [Perdew-Burke-Ernzerhof (PBE) functional \cite{Perdew96PBE}]. GTH pseudopotentials \cite{Goedecker96, Hartwigsen98} were used to replace the sharp
core electron density, with corresponding GTH-DZVP ($2s2p3s3p3d$ AOs for B and N, and $3s3p3d4s4p$ AOs for Si) and GTH-DZVP-MOLOPT-SR ($3s3p3d4s4p4d4f5s$ AOs for Ni, and $2s2p3s3p3d$ AOs for O) basis sets~\cite{VandeVondele2007} used to
represent the valence electrons. 
Gaussian density fitting was used to compute the two-electron integrals~\cite{Sun17}. \ZHC{We used an even-tempered Gaussian basis \cite{Stoychev17ETBbasis} as the density fitting auxiliary basis, i.e. $L_{nl} (r) \propto r^l \exp(\alpha \beta^n r^2)$, where we used the exponential factor $\beta = 2.3$ for NiO and $\beta = 2.0$ for all other systems. The number of fitting functions was chosen to ensure high accuracy, and thus the size of the auxiliary basis is about 10 times as large as the number of AOs.} The GTH-SZV (h-BN and Si) and GTH-SZV-MOLOPT-SR (NiO) basis functions were used as the reference free-atom AOs 
to construct the IAOs. 
In the mean-field calculations used to derive the embedding Hamiltonian and in the DMET self-consistency,
we sampled the Brillouin zone with a $\Gamma$ centered mesh chosen so as to be able to fit unit multiples of the DMET impurity supercell.
These included a $6\times 6 \times 1$ mesh for BN, and a $4 \times 4 \times 4$ mesh for Si and NiO. Larger meshes were used in independent
estimates of the mean-field TDL for BN (up to $12 \times 12 \times 1$) and Si (up to $8\times 8 \times 8$).
All mean-field calculations were converged to an accuracy of better than $10^{-10}$ a.u. per unit cell. In the case
of Hartree-Fock energies, all energies included the leading-order exchange finite-size correction (probe-charge Ewald~\cite{Paier05, Sundararaman13regularization}, \texttt{exxdiv=ewald} in \textsc{PySCF}). Note that the above correction applies to all DMET energies as these
use the Hartree-Fock expression for the mean-field energy even when density functional orbitals are used. 

{\bf{Impurity solver.}} We used coupled cluster singles and doubles (CCSD) \cite{Bartlett07} as an impurity solver, as
implemented in \textsc{PySCF} \cite{Sun18pyscf}, which is able to treat a large number of orbitals efficiently. In NiO where DMET self-consistency produced symmetry breaking, we used unrestricted CCSD (UCCSD).
The CC density matrices were obtained from the CC $\Lambda$ equations~\cite{Shavitt09book}. The CC energies were converged to $10^{-8}$ a.u.. 

{\bf{DMET self-consistency.}} For BN and NiO, the correlation potential $u$ was added to only the valence orbitals and for Si,
$u$ was added to all impurity orbitals as this gave smoother DMET convergence. We carried out CSC calculations for all three systems, and included additional non-CSC results of NiO for comparison.  
The convergence criterion on the DMET self-consistency was chosen such that the maximal change of an element in $u$ was less than $5\times 10^{-5}$ a.u., which corresponded roughly to an energy accuracy of better than $1\times 10^{-5}$ a.u..

\section{Results and Discussion}\label{sec:results}

\subsection{2D Boron Nitride}\label{subsec:BN}

We first study the behavior of DMET on a 2D boron nitride monolayer. In a GTH-DZVP basis, BN has a unit cell of 2 atoms, with $2s2p$ AOs on
each atom giving 8 valence orbitals per cell, and $3s3p3d$ AOs on each atom providing 18 higher-energy virtual orbitals per cell.
We illustrate the valence IAOs of boron in BN in Fig. \ref{fig:orb iao}. 
\begin{figure}[hbt]
\subfigure[]{\label{fig:orb
	iao}\includegraphics[width=0.4\textwidth]{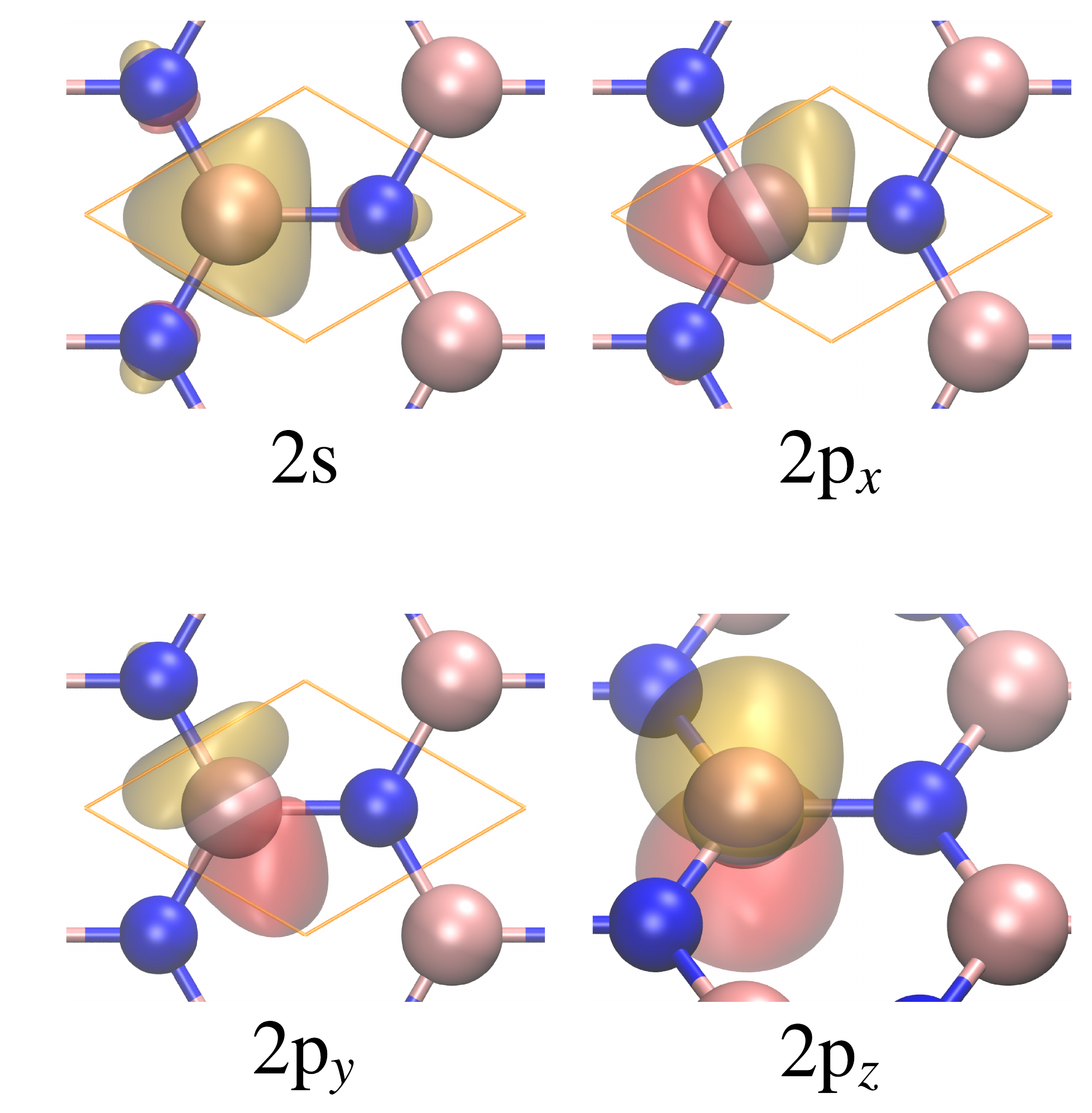}}
\subfigure[]{\label{fig:orb bath}\includegraphics[width=0.4\textwidth]{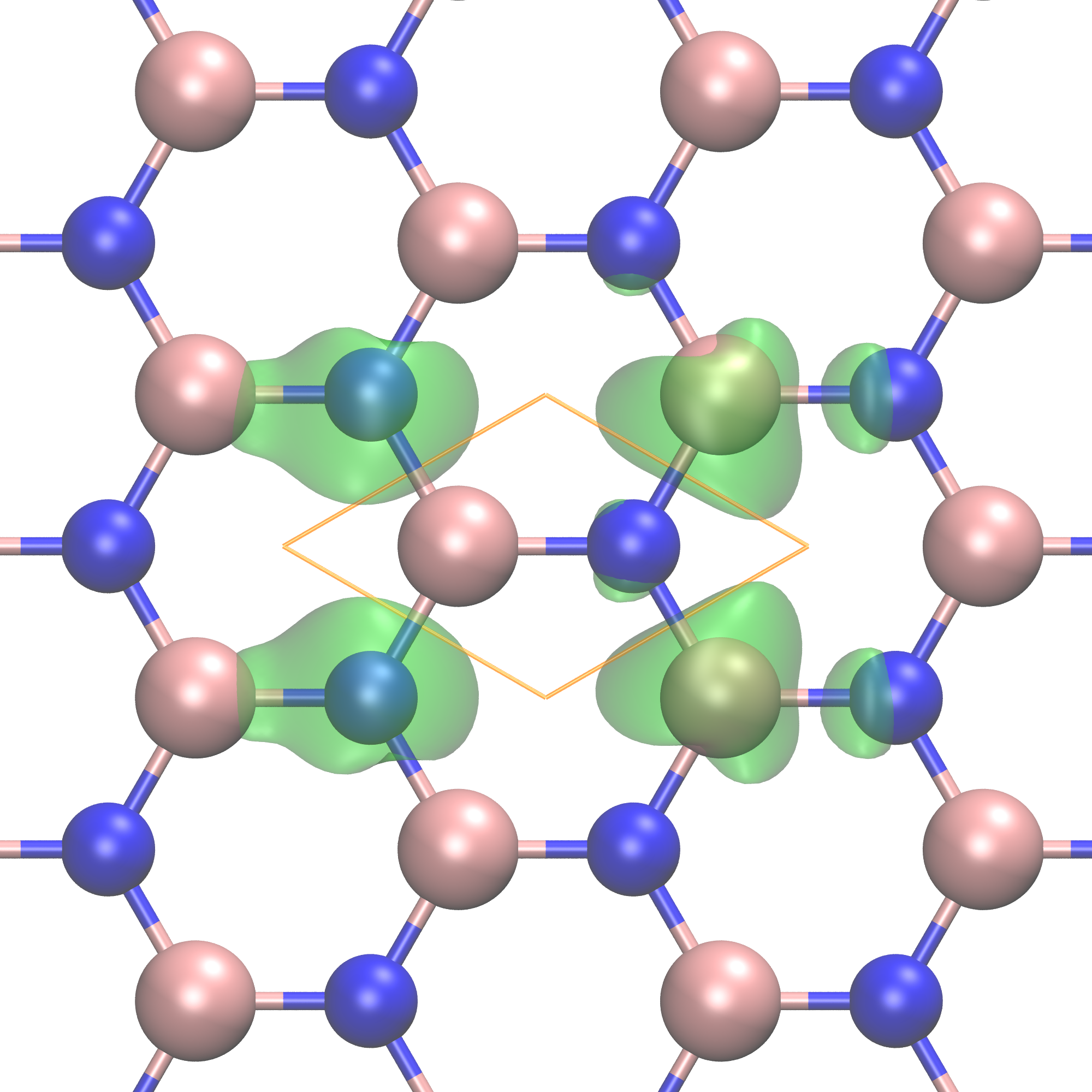}}
\caption{Impurity orbitals and bath density of BN used in the DMET calculations. The boron and nitrogen atoms are colored pink and blue respectively. (\lroman{1}) Impurity valence orbitals associated with one boron atom (IAOs from boron). (\lroman{2}) Bath orbital density coupled to the first reference cell.}
\end{figure}
As expected, the IAOs of boron are quite local, retaining their original AO character but with some slight polarization to reflect
the mean-field solution in the crystal environment. The bath orbital density is plotted in Fig. \ref{fig:orb bath} (we only show the total density summed
over the bath orbitals here, since the embedded problem only depends on the linear span of the bath). It is clear that the bath orbitals are localized around the impurity cluster and give an effective representation of the remainder of the boron nitride crystal. In particular, the bath orbitals
serve to terminate the dangling bonds on the impurity boundary, thus turning the embedding problem into a closed-shell one at the mean-field level.
The impurity valence orbitals and bath orbitals pictured here, together with the impurity virtual orbitals (not shown), constitute
the embedding orbitals.

\begin{figure}[hbt]
\includegraphics[width=0.5\textwidth]{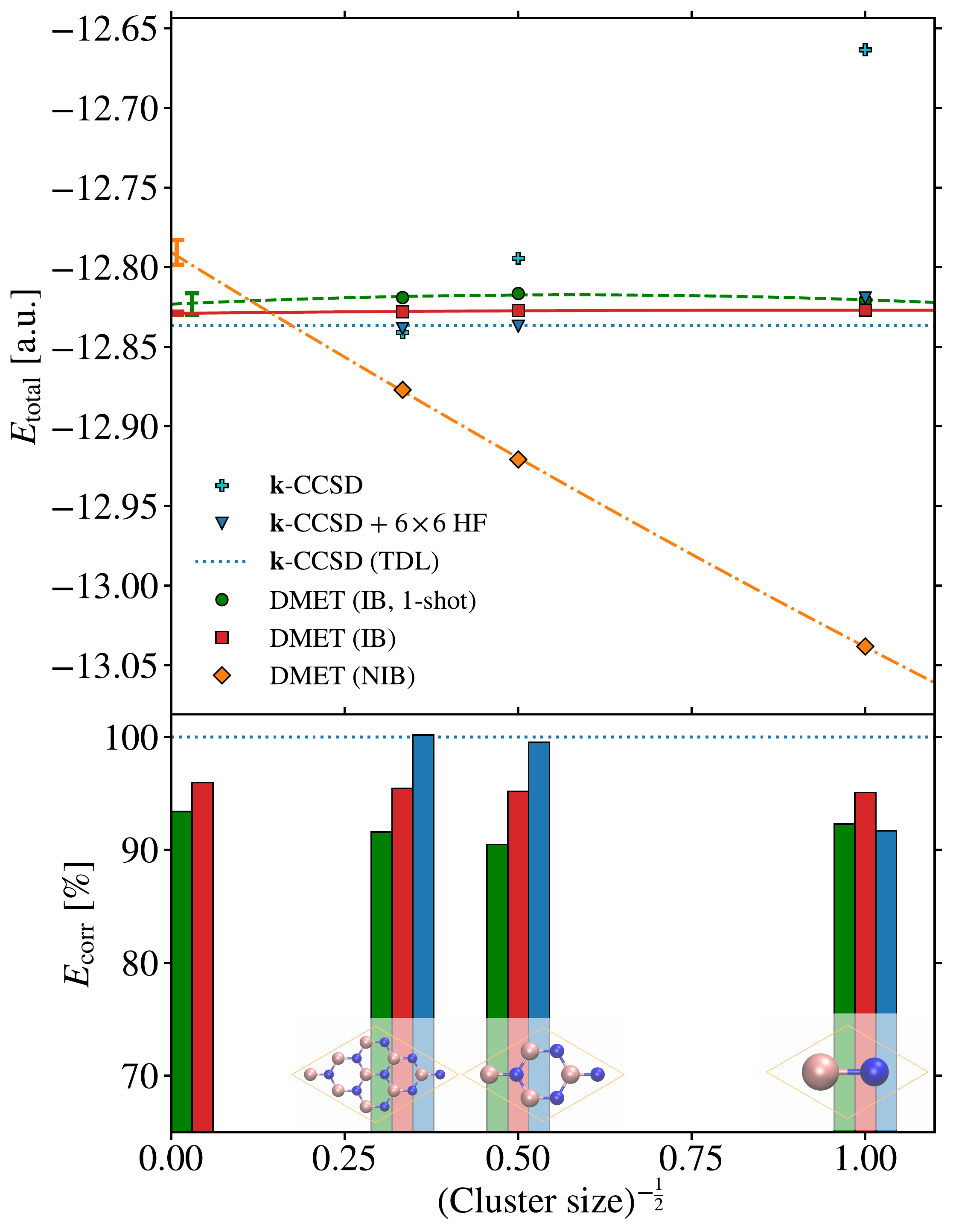}
\caption{Upper panel: Total energy from DMET compared with $\veck$-sampled CCSD. In the case of DMET with the interacting bath (IB), both one-shot and self-consistent energies are reported. DMET with non-interacting bath (NIB) is also shown for comparison. The extrapolated values of DMET is from an average of linear regression and quadratic fitting. The error bar is the difference
between the linear and quadratic fitted values. \ZHC{ We plot the energy of $\veck$-CCSD with small $\veck$-mesh (one curve with HF energy at corresponding small $\veck$-mesh and the other with HF energy at $6\times 6$ $\veck$-mesh) and the extrapolated TDL results as reference. Lower panel: Correlation energy ratio with respect to the extrapolated CCSD correlation energy.} }\label{fig:BN extrapolation}
\end{figure}

We computed total energies (per cell) from DMET for different cluster sizes, $1\times 1$, $2\times 2 $ and  $3\times 3 $. We compare
these total energies to those from $\veck$-sampled periodic CCSD ($\veck$-CCSD)
extrapolated to the TDL (see Fig. \ref{fig:BN extrapolation}) \ZHC{which has recently been
  demonstrated to be a high accuracy method in a variety of
  different materials~\cite{McClain17, Gao19CCTMO, Zhang19KCCreview}.
Note that,  accounting fully for the $\veck$-point symmetry, 
$\veck$-CCSD has a computational scaling of $n_{\AO}^6 n^4_{\veck}$.}
The reference TDL $\veck$-CCSD energy is the sum of the extrapolated HF energy using a large $\veck$-mesh (up to $12\times 12 \times 1$, extrapolating
with the form $n^{-1}_{\veck}$ after using the Ewald exchange divergence correction~\cite{Gygi86, Paier05}) and the extrapolated
$\veck$-CCSD correlation energy using a smaller $\veck$-mesh (up to $6\times 6 \times 1$, extrapolating with the form $n^{-1}_{\veck}$). 
Compared to the TDL reference energy, even using the smallest ($1 \times 1$) cluster, DMET gives an accurate total energy that captures about 95\% of the correlation energy. Extrapolating over the DMET cluster size (using the surface to volume form $N_{\mathrm{c}}^{-1/2}$, where $N_{\mathrm{c}}$ is the cluster size)
further improves the accuracy by about 1-2\% in the correlation energy. The one-shot DMET result (i.e. without DMET self-consistency) is less accurate than the self-consistent one by $\sim 8$ mHartree (3\% of the correlation energy), demonstrating the
contribution of self-consistent matching between the high-level calculation and the low-level mean-field calculation.
We note that  self-consistency is generally not very important in non-magnetic weakly-correlated systems, as there
are no symmetry broken phases to be generated by DMET, and only provides a modest quantitative correction to the observables.

Compared to small $N\times N\times 1$ $\veck$-mesh CCSD energies, the DMET total energies are more accurate
for the $1 \times 1$ and $2 \times 2$ cluster sizes, but less accurate for the $3 \times 3$ case.
\ZHC{The finite size 
 error in the total
energy, arising from the finite $\veck$-mesh or DMET cluster size, can be separated into two sources, (\lroman{1}) the finite size error in the mean-field energy and (\lroman{2})
the finite size error in the many-body correlation energy.} For embedding methods like DMET, the error from the first
source is (largely) eliminated. Thus, as shown in Fig. \ref{fig:BN extrapolation}, the DMET total energy is good even for
a small cluster size. In the CCSD calculation, however, the error from (\lroman{1}) is large for small clusters, and
therefore, a potentially better recipe for the total energy is to sum the HF energy from a larger cluster (or even extrapolated
to the TDL) and the correlation energy from the small cluster calculation. \ZHC{In the upper panel, we show the $\veck$-CCSD correlation energy added to the $6\times 6$ HF energy (corresponding to the size of the DMET lattice), as well as to
the  extrapolated TDL HF energy.}
Together with the data
in the lower panel of Fig. \ref{fig:BN extrapolation}, we see that the correlation energy
$E_{\mathrm{corr}}$ of CCSD, which relies on the above error cancellation, is already very accurate for the $2
\times 2$ cluster and is better than that of DMET for this cluster size. It is then worth analyzing the source of errors in the
small cluster DMET correlation energy. One
source is the lack of embedding of the non-valence virtual orbitals, which are localized to the reference cell with the periodicity
of the large DMET mean-field lattice, not the periodicity of the impurity (as in the $\veck$-CCSD calculation). The advantages of DMET
in the current implementation thus manifest when the predominant correlation is within the valence space itself (which is fully embedded) as is
typical of strong correlations, rather than primarily involving excitations to non-embedded, non-valence, virtual orbitals as in this system.
One way to diminish the boundary effect on the DMET non-valence virtuals 
is to evaluate the energy from the central part of the supercell, for which the surrounding atoms effectively provide a bath for the virtuals.
We find then that the energy evaluated using the central cell of the embedded cluster
covers 103.8\% of the correlation energy (using the preceding $3\times 3$ cluster calculation) or
100.1\% (if no chemical potential fitting is used), which is better than that obtained by direct energy evaluation using
the entire embedded cluster. It may be possible to further reduce this boundary error using the dynamical cluster approximation formulation of DMET (DCA-DMET)\cite{Zheng17} or bootstrap embedding\cite{Welborn16bootstrap, Ricke17bootstrap, Ye19bootstrap}. 



We finally consider DMET results obtained using the non-interacting bath (NIB), as also shown in Fig. \ref{fig:BN extrapolation}. We see that although the extrapolation is quite systematic, the accuracy is worse than that of the interacting bath for all three cluster sizes.
This result is generally found in chemical systems with long-range Coulomb interactions, as the interacting bath carries some
information about the inter-cluster interactions. However, the NIB formalism has the potential computational advantage that 
the construction of the NIB embedded Hamiltonian is cheaper than the IB one, since only the impurity part of the two-particle Hamiltonian is needed. In addition, the correlation potential can be used to mimic the effect of the long-range Coulomb contributions to the Fock matrix.
This makes the NIB scheme an interesting possibility in large systems.



\subsection{Bulk Silicon}\label{subsec:Si}

We next test the ability of DMET to describe the structural properties of bulk Si. We performed a series of calculations on different primitive cell volumes and fitted the relative total energy $E$ as a function of the volume $V$ using the Birch-Murnaghan (B-M) equation of state (EOS) \cite{Murnaghan44, Birch47}, from which the equilibrium volume and bulk modulus can then be determined. To obtain accurate results for the TDL, we considered three clusters of different shapes: a $1\times 1 \times 1$ primitive cell (2 Si atoms), a conventional diamond cubic cell (8 Si atoms) and a $2\times 2 \times 2$ supercell (16 Si atoms). We performed the extrapolation with respect to cluster volume $V_{\mathrm{c}}$ using
\begin{equation}\label{eq:extrapolation Si}
E(V_{\mathrm{c}}) = E(\infty) + a_0 V^{-1/3}_{\mathrm{c}} + \cdots 
\end{equation}
The total energy includes the correction from HF at the TDL. The equilibrium volumes and bulk moduli are collected in Table \ref{tab:bulk property Si}.

\begin{figure}[hbt]
\includegraphics[width=0.5\textwidth]{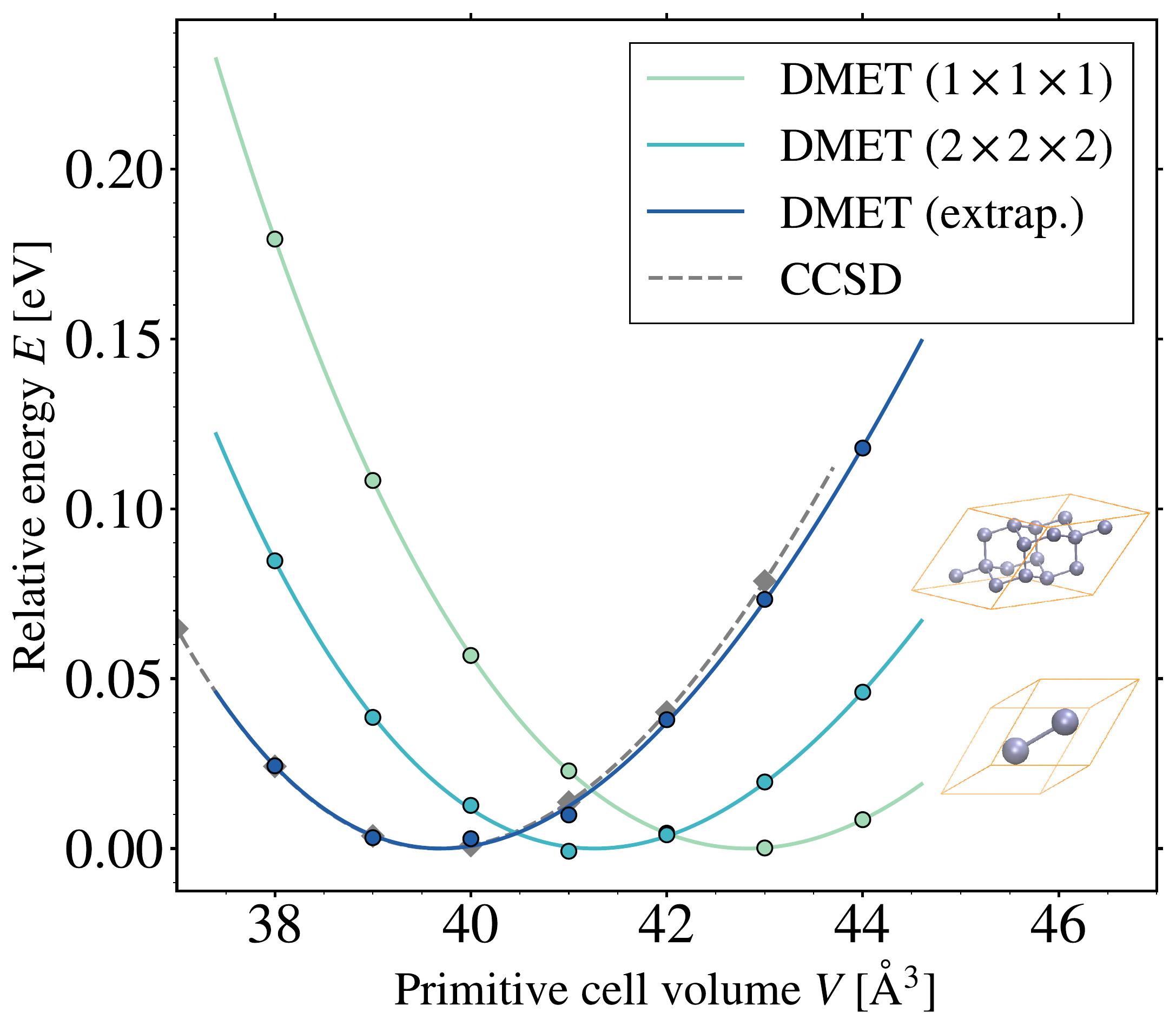}
\caption{Equation of state curves of Si from DMET and CCSD. For DMET, we omit the cubic cell curve for clarity. CCSD data is taken from Ref. \onlinecite{McClain17}. }\label{fig:eos Si}
\end{figure}

\begin{table}[ht!]
	\centering
	\caption{Equilibrium volume of the primitive cell $V_0$ and bulk modulus $B_0$ of silicon from different approaches.
    The extrapolated values are from the linear fit of $1\times 1\times 1$ and $2\times 2 \times 2$ results. The CCSD results are taken from Ref. \onlinecite{McClain17} \ZHC{, which uses the larger GTH-TZVP basis}. The experimental $V_0$ is from Ref. \onlinecite{Tobbens01Si} and $B_0$ is from Ref. \onlinecite{Schimka11} with a zero-point correction.}
	\label{tab:bulk property Si}
	\begin{tabular}{lcccc}
	\hline\hline
Methods	 &     & $V_0$ [$\text{\AA}^3$] &  $B_0$ [GPa]  \\
	\hline
HF   & extrap.                &  40.30      &  107    \\
DMET & $1 \times 1 \times 1$  &  42.83      &  87.9   \\
     & cubic cell             &  41.90      &  88.5   \\
     & $2 \times 2 \times 2$  &  41.26      &  91.1   \\
     & extrap.                &  39.69      &  99.0    \\
CCSD & $3 \times 3 \times 3$  &  39.21      &  103    \\
Expt.&                        &  40.04      &  101    \\
    \hline\hline
	\end{tabular}
\end{table}
From the table, we see that the equilibrium volume of DMET using the $1\times 1 \times 1$ cluster deviates from the experimental value by 7\%.
The error from the smallest impurity cluster is thus larger for Si than for BN. This is because Si has a much smaller band gap and thus less local
correlation involving the non-valence space. However, the results improve rapidly when increasing the size of cluster. To illustrate this, we show the EOS curves for different cluster sizes in Fig. \ref{fig:eos Si}.
It is clear that the $1\times 1\times 1$ curve is shifted to larger volume compared to experiment or CCSD. Increasing the cluster size
systematically shifts the curve back towards experiment and the $\veck$-CCSD benchmark, resulting in a very small relative error (w.r.t. experiment) of 0.9\% for $V_0$ for the extrapolated curve.
The extrapolated bulk modulus $B_0$ also agrees well with the experimental and $\veck$-CCSD benchmark values. \ZHC{Overall, the accuracy achieved by extrapolated DMET appears comparable to that of the $\veck$-CCSD benchmark in a full $3 \times 3 \times 3$ periodic calculation, although we note that a different basis was used.}


\subsection{Nickel monoxide}\label{subsec:NiO}

We now demonstrate the ability of DMET to treat a more strongly correlated problem by considering a typical transition metal compound, NiO.
Below the N\'eel temperature, NiO displays an antiferromagnetic (AFM) phase with a staggered magnetization along the [111] direction (the so-called
AFM-\uroman{2} phase). Although DFT (with PBE) and HF do predict spin-polarization, it is known that DFT often underpolarizes
while HF often overpolarizes antiferromagnetic states. To avoid such biases in the DMET calculation,
we embed the DMET calculation in an initial \emph{unpolarized} mean-field state. 
We constructed
the unpolarized mean-field state
by using the orbitals
obtained from the spin-averaged Fock matrix of an unrestricted Hartree-Fock or DFT calculation.
We use the spin-averaged Fock matrix for convenience because
without finite-temperature smearing, the restricted calculations either have difficulty converging due to the
metallic nature (DFT) or exhibit an unphysical symmetry breaking of the density between the symmetry-equivalent nickel
atoms (HF).
The spin-averaged Fock matrix is similar to the restricted one with smearing but exactly preserves the symmetry between the two nickel atoms. We denote 
DMET calculations based on the spin-averaged mean-field orbitals by $\DMET@\Phi_{\RHF}^{*}$ ($\DMET@\Phi_{\RPBE}^{*}$), where ``$*$''
means the restricted orbitals are actually from the spin-averaged unrestricted Fock matrix rather than a real restricted one.

The spectrum of such a spin-averaged Fock matrix is gapless. After adding an initial DMET correlation potential, e.g.
taken from the local part of the UHF polarized potential, the system becomes gapped and  $S^2$ symmetry is broken. Without
CSC, the final DMET mean-field gap is $\sim 3$ eV and with CSC, the DMET mean-field gap is $\sim 10$ eV, closer to the Hartree-Fock mean-field gap ($\sim 12$ eV). \ZHC{(Note that the experimental band gap of AFM NiO is $\sim 4.3$ eV~\cite{Sawatzky84}).} It should be emphasized that although the band gap from the DMET lattice mean-field reflects the insulating nature of the system, its value does not correspond to the true fundamental gap of the system. Even if the density from the impurity solver were exact and the matching between density matrices were perfect, the mean-field gap is not exact due to the derivative discontinuity contribution\cite{Perdew17}, similar to the Kohn-Sham gap obtained from an optimized effective potential (OEP) calculation \cite{Kuemmel08RMP}. 
  

The ground state charges and local magnetic moments of NiO from DMET starting from different initial mean-fields (spin-averaged HF and PBE) are summarized in  Table \ref{tab:charge and magnetic moment}. Assignment of local observables to different atoms (population analysis)
was performed using the IAOs + PAOs and the density matrix from the CC impurity solver.
\begin{table}[hbt]
	\centering
	\caption{Local charge (in $e$) and magnetic moment (in $\mu_{\mathrm{B}}$) of NiO from different methods. The values on Ni (O) are averaged from the two Ni (O) sites in the primitive cell. We include the DMET results from different initial orbitals ($\Phi_{\RHF}^{*}$ and $\Phi_{\RPBE}^{*}$), with / without charge self-consistency (CSC).
The experimental data is taken from Refs. \onlinecite{Alperin62NiOm, Fender68NiOm, Cheetham83NiO}. }
	\label{tab:charge and magnetic moment}
	\begin{tabular}{lcccc}
	\hline\hline
Methods	   & $\rho_{\mathrm{Ni}}$    &  $m_{\mathrm{Ni}}$ &  $m_{\mathrm{O}}$  \\
	\hline
HF               &  1.42    &  1.86    &  0.000    \\
PBE              &  1.02    &  1.42    &  0.000    \\ 
$\DMET@\Phi_{\RHF}^{*}$ w/o CSC  &  1.32    &  1.77    &  0.018    \\ 
$\DMET@\Phi_{\RPBE}^{*}$ w/o CSC &  1.27    &  1.74    &  0.017    \\
$\DMET@\Phi_{\RHF}^{*}$ w/ CSC          &  1.37    &  1.81    &  0.001    \\ 
$\DMET@\Phi_{\RPBE}^{*}$ w/ CSC        &  1.35    &  1.78    &  0.000    \\
Expt.            &          &  1.70-1.90 &           \\
    \hline\hline
	\end{tabular}
\end{table}
We also include unrestricted HF, PBE results for comparison.

First, we observe clear charge transfer from Ni to O in all methods. Among them, HF gives the largest ionic character
while PBE smears out the charge and predicts the smallest charge transfer. The DMET results from different starting
orbitals and CSC conditions are between these two limits and are relatively close to each other. The DMET results with
CSC (starting from HF and PBE) are particularly close to each other as the inter-cluster part of density matrix is updated using
information from the high-level embedded calculation. In fact, in the case of CSC, the only effect of the initial choice of orbitals in DMET
on the final result comes from the different definition of the local orbitals.

\begin{figure}[hbt]
\includegraphics[width=0.5\textwidth]{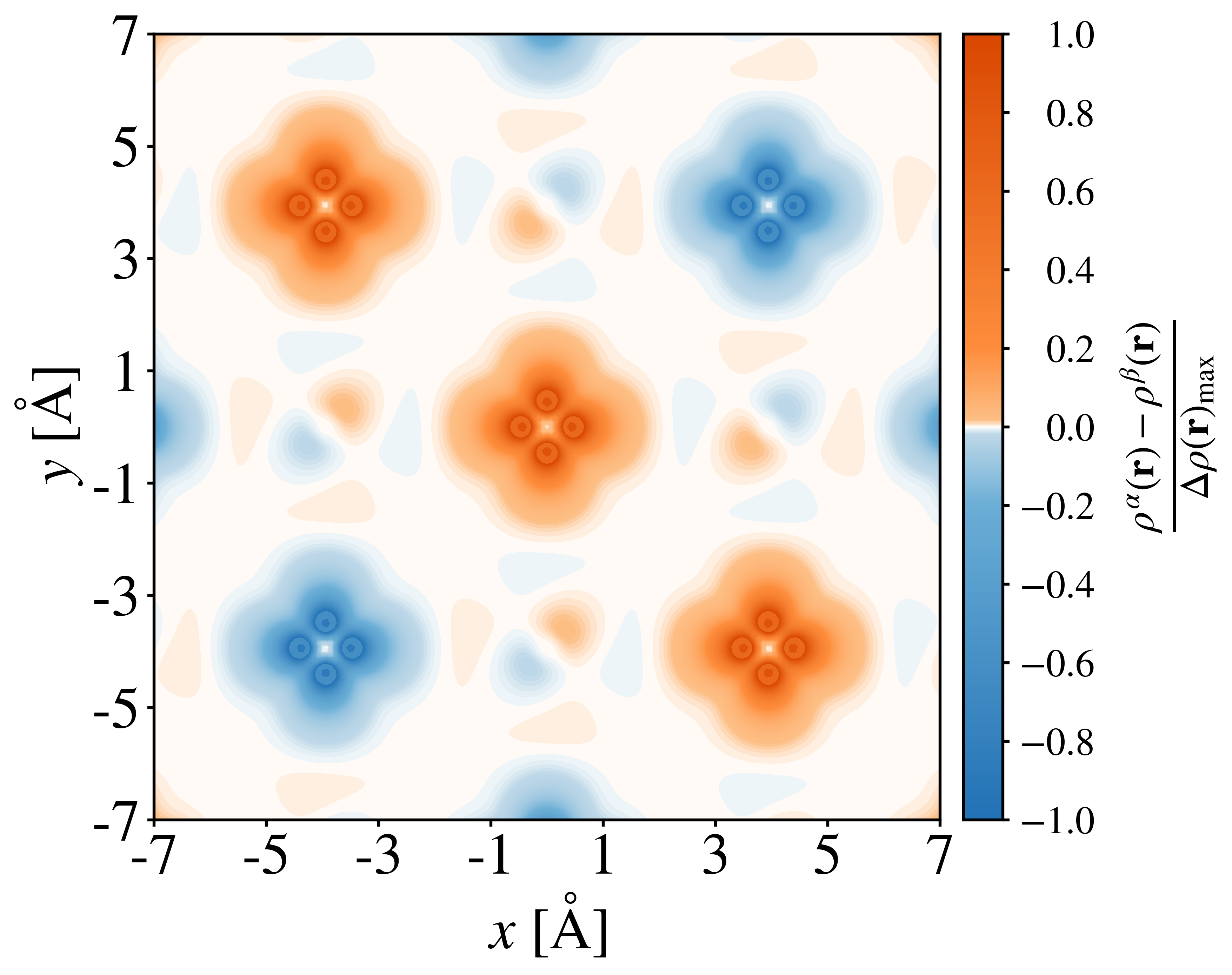}
\caption{Spin density $\rho_{\alpha} - \rho_{\beta}$ on the  (001) plane of NiO from $\DMET@\Phi_{\RHF}^{*}$ with charge self-consistency.}\label{fig:NiO spin density}
\end{figure}

Compared to the experimental estimate of the magnetic moment, unrestricted Hartree-Fock gives a Ni magnetic moment
at the higher-end of the experimental range, while PBE severely underestimates the magnetic moment. DMET
yields results independent of the starting orbitals with a moment that agrees well with experiment.
To illustrate the AFM distribution in NiO, we plot the spin density distribution in the (001) plane of NiO in Fig. \ref{fig:NiO spin density}.
In the figure, the $\alpha$- and $\beta$- spin planes alternately appear along the diagonal direction, showing a clear AFM pattern.
In particular, the spin density on Ni is in the shape of the $d_{x^2-y^2}$ orbital, indicating that its occupation is
asymmetric with respect to the $\alpha$ and $\beta$ electrons. In fact, the $t_{2g}$ orbitals are almost fully occupied ($\sim 5.97$ $e$ in our population analysis), and the $e_g$ orbitals ($d_{x^2-y^2}$ and $d_{z^2}$) are occupied only in one spin sector ($\sim 1.99$ $e$), and roughly empty in the other ($\sim 0.19$ $e$).  The local magnetic moment on Ni therefore mainly comes from the contribution of the $e_g$ electron density, as expected
from crystal field theory. The density on oxygen is in the shape of a $p$ orbital and is polarized according to its orientation relative to Ni. The average polarization on oxygen should be close to zero due to symmetry. As shown in Table \ref{tab:charge and magnetic moment}, the magnetic moments on oxygen from DMET (especially with CSC) are indeed close to zero.

We now take a closer look at the spin-spin correlation in NiO. To this end, we evaluate the spin-spin correlation function
between the two nickels in the unit cell, 
\begin{equation}\label{eq:spin-spin corr func NiO}
\sum_{i\in \ce{Ni1}, j\in \ce{Ni2}}\expval{\vecS_i \cdot \vecS_j} = \sum_{i\in \ce{Ni1}, j\in \ce{Ni2}} \sum_{a=x, y, z}\expval{S_i^{a}S_j^{a}},
\end{equation}
where $i$ and $j$ are the indices of LOs located on the first and second Ni respectively. In the 
DMET@$\Phi^*_{\mathrm{RHF}}$ calculation with charge self-consistency, the expectation value is $-0.8147$, where the minus sign
arises from the AFM correlation between the spins of two nickels. This value, however, is very close to the product 
$\expval{S^{z}}\expval{S^{z}}=-0.8149$. 
In addition, the spin non-collinear contributions ($\expval{S^{x} S^{x}}$ and  $\expval{S^{y}
  S^{y}}$) are almost zero (note that the calculation spontaneously chooses a $z$ magnetization axis due to the initial unrestricted
Hartree-Fock reference or form of the correlation potential). All these features suggest that the ground-state of the AFM spin lattice
in NiO is close to that of a classical Ising model, rather than a quantum one. Our results are consistent with experimental measurements on the critical behavior of the magnetic phase transition in NiO \cite{Chatterji09, Germann74, Negovetic73}, where the critical exponents are found to be 
very close to those of the 3D Ising model.
 
\begin{figure}[hbt]
\includegraphics[width=0.5\textwidth]{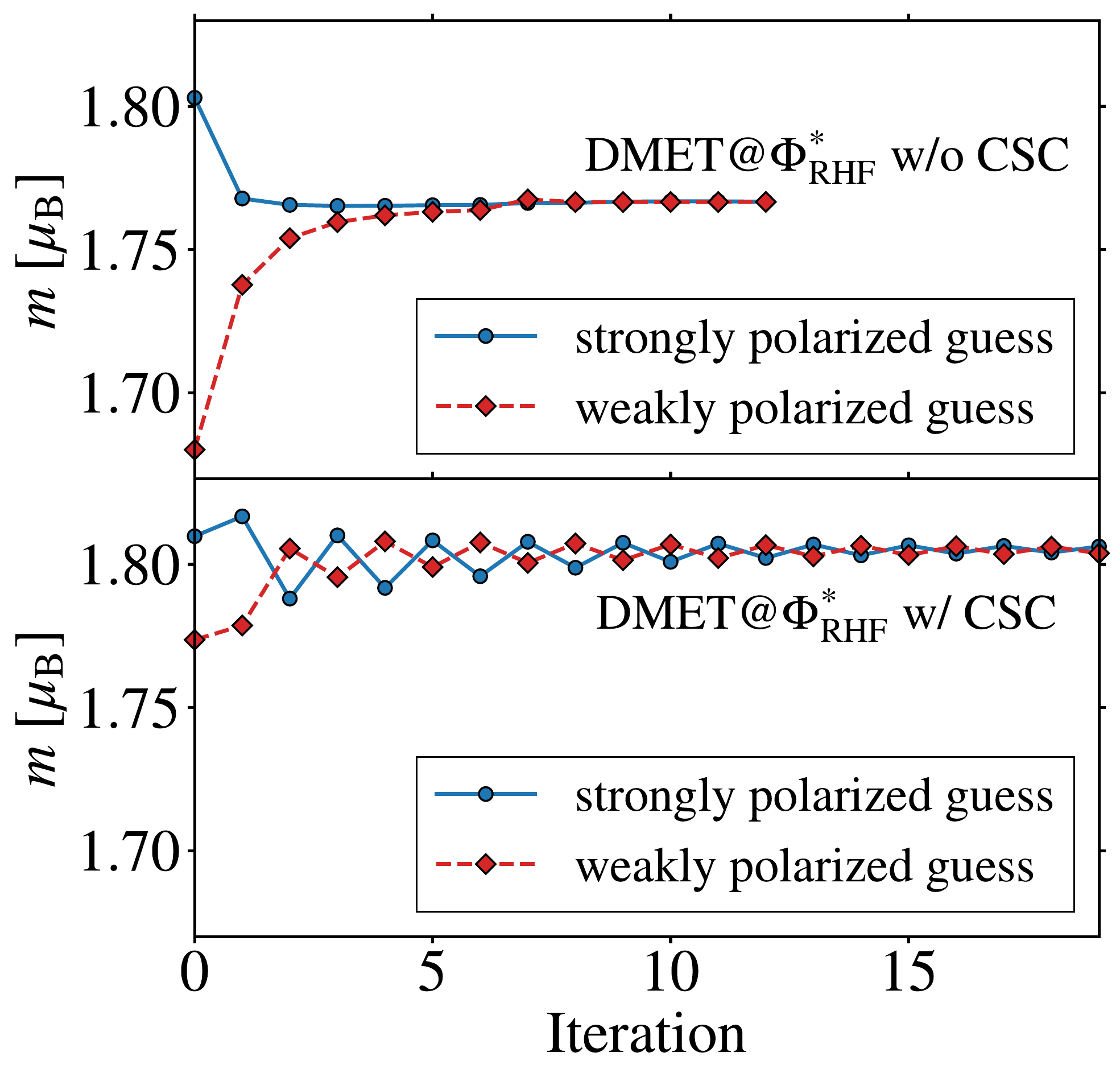}
\caption{The convergence of the magnetic moment on Ni from different initial correlation potentials. Upper panel: $\DMET@\Phi_{\RHF}^{*}$ without CSC using different initial guesses: UHF potential (strongly polarized) or UHF potential scaled by 0.1 (weakly polarized). Lower panel: The same as the upper panel but with CSC.}\label{fig:m-vs-iter}
\end{figure}

In the above results, we found that the DMET order parameters are insensitive to the initial mean-field orbitals,
due to the DMET self-consistency. As discussed in section~\ref{subsec:dmet}, this self-consistency 
contains two different contributions: self-consistency of the DMET correlation potential (expressed along the cluster blocks
of the mean-field lattice Hamiltonian) and charge self-consistency of the mean-field Fock operator (for the off-diagonal
blocks of the mean-field lattice Hamiltonian). To show the robustness of the self-consistency with respect to the
correlation potential guess and the relative magnitude of these two contributions, 
we show the convergence of the local magnetic moment of Ni with respect to the number of iterations in Fig. \ref{fig:m-vs-iter} (for
initial restricted orbitals from a spin-averaged Fock matrix $\Phi_{\RHF}^{*}$) with
two different initial guesses for the correlation potential: the strongly polarized UHF potential, and a weakly polarized potential
equal to the UHF potential scaled by a factor 0.1, both with and without charge self-consistency.
From the figure, we see that starting from different initial guesses for the  correlation potential, the magnetic moments from non-self-consistent
(i.e. one-shot) DMET (the \nth{0} iteration in Fig. \ref{fig:m-vs-iter}) can be very different. However, after only 1 step, the magnetic moments are
significantly improved. Eventually, the magnetic moments from the two guesses converge to a very similar value, showing that the DMET self-consistency effectively removes the initial correlation potential guess dependence. The picture with and without charge self-consistency is very similar, showing
that the DMET correlation potential is the main factor controlling the local order parameter. Note that in Fig. \ref{fig:m-vs-iter}, the LOs are the same (based on Hartree-Fock) for all calculations and hence there is no initial LO dependence. \ZHC{Finally, as a rough indicator
  of cost, each DMET iteration takes about 1 hour (the computational setup is described in Sec.~\ref{sec:compdetails}).}


\section{Conclusions}\label{sec:conclusion}

In this paper, we described an \abinitio quantum embedding scheme for density matrix embedding calculations in solids,
focusing on the practical implementation choices needed for an efficient computational scheme. Our tests on the BN, Si, and NiO systems, that
span a range of electronic structure, demonstrate that our implementation can handle both realistic unit cells and basis sets.
The strengths of DMET are most visible in the simulations of NiO, where the wide spread in magnetic behavior generated by
different mean-field approximations is almost entirely removed in the subsequent DMET calculation. In more weakly correlated systems, more work
is needed to improve the quantitative accuracy of DMET arising from the treatment of excitations to non-valence orbitals,
which are not fully embedded in our scheme.
Overall, however, our results lead us to be optimistic that this computational framework provides a means
to realize \abinitio calculations on interesting correlated solids using density matrix embedding theory. Much of the computational
framework can be reused also to realize \abinitio dynamical mean-field theory (DMFT) in solids, and elsewhere, we report the results of such a scheme.\cite{Zhu19-dmft-solid}.



\begin{acknowledgement}
  We thank James McClain for providing CCSD data on the equation of state of Si, Lin Lin and Yang Gao for helpful discussions and
    Mario Motta for helpful comments on the manuscript. This work is partially supported by
  US Department of Energy via award no. DE-SC19390. Additional support was provided by the Simons Foundation via an Investigatorship and through
  the Simons Collaboration on the Many-Electron Problem.
\end{acknowledgement}

\begin{appendix}

\section{k-adapted IAO and PAO}\label{app: kiao}

The key ingredients for IAO construction \cite{Knizia13IAO} are the occupied MOs $\qty{\ket{\psi_{m}}}$ and two sets of
    bases, $B_1$ and
    $B_2$. Concretely, $B_1$ is the normal AO basis used in the mean-field calculation (labeled by $\mu, \nu, \cdots$)
    and $B_2$ is the reference minimal basis set (labeled by $\rho, \sigma, \cdots$). $B_1$ usually contains the space
    of $B_2$ and the extra part reflects the polarization. The goal of IAO construction is to obtain a set of AO-like
    orbitals that contains the occupied space but has the size of the small basis set $B_2$. To achieve this, we first define the \emph{depolarized} MOs $\qty{\ket{\psi_{\bar{m}}}}$ by projecting the MOs to $B_2$, then back to $B_1$,
\begin{equation}\label{eq: IAO depolarized MO}
\ket{\psi_{\bar{m}}} = \mathrm{orth}\qty(P^{B_1} P^{B_2} \ket{\psi_m}),
\end{equation}
where $P$ is the resolution of identity (or projector) of AOs, e.g.
\begin{equation}\label{eq: IAO AO projector}
P^{B_1}_{\mu \nu} = \sum_{\mu \nu} \ket{\phi_{\mu}} S^{B_1}_{\mu \nu} \bra{\phi_{\nu}}.
\end{equation}

Using the depolarized MO projector $\bar{O} \equiv \sum_{\bar{m}}\dyad{\psi_{\bar{m}}}{\psi_{\bar{m}}}$, we can split the $B_2$ set into occupied ($\bar{O} \ket{\phi_{\rho}}$) and virtual spaces $\qty(1-\bar{O}) \ket{\phi_{\rho}}$. The IAOs $\qty{\ket{w_i}}$ are obtained by further projecting these two subspace bases onto their polarized counterparts ($O \equiv \sum_{m}\dyad{\psi_m}{\psi_m}$ and $1-O$) and applying L{\"o}wdin orthogonalization,
\begin{equation}\label{eq: IAO expression}
\ket{w_i} = \mathrm{orth}\qty{\qty[O \bar{O} + \qty(1-O) \qty(1-\bar{O})] \ket{\phi_\rho}}.
\end{equation}

In periodic systems, the quantities in the above equations should be understood to carry $\veck$ labels, e.g.
$\ket{\phi_{\mu}} \rightarrow \ket{\phi_{\mu}^{\veck}}$ is a crystal AO, and $S^{B_1} \rightarrow S^{\veck, B_1}$ is the corresponding overlap matrix. 
These quantities are already evaluated in the mean-field calculations. The only thing we need additionally is the overlap matrix between basis $B_1$ and $B_2$, which can be evaluated directly,
\begin{equation}\label{eq: IAO S12}
S^{\veck, B_1, B_2}_{\mu \rho} = \int \diff\vecr \sum_{\vecT} \ee^{\ii \veck\cdot \vecT} \phi_{\mu}^{*} (\vecr)  \phi_{\rho} (\vecr - \vecT),
\end{equation}
where the summation is over the periodic images $\vecT$. After the IAOs are constructed, the $\veck$-adapted PAOs are obtained by projecting out the IAO components from the AOs at each $\veck$-point.

\section{Embedding ERI construction with FFTDF}\label{app:fftdf eri}

The embedding ERIs can also be constructed from FFTDF, which uses the fast Fourier transform to represent the Coulomb kernel and to expand the AO pairs. In such a case, $L$ in Eq. \ref{eq:density fitting} is a set of planewaves $\qty{\vecG}$ \cite{McClain17},
\begin{equation}\label{eq:FFTDF}
\eri{\mu \veck_\mu \nu \veck_\nu}{\kappa \veck_\kappa \lambda \veck_\lambda} \approx \Omega^2 \sum_{\vecG} \eri{\mu\veck_\mu \nu \veck_\nu}{\vecG} \frac{4\pi}{\Omega\qty|\vecq + \vecG|^2} \eri{-\vecG}{\kappa \veck_\kappa \lambda \veck_\lambda} ,
\end{equation}
where $\Omega$ is the volume of the unit cell, $\vecq \equiv \veck_\mu - \veck_\nu$ and only three $\veck$s are independent. Similarly to the algorithm for GDF, the AO-to-EO transformation can be performed on the 3-index quantities. The procedure is described in Algorithm \ref{alg:eri with fftdf}. 
\begin{algorithm}[H]
  \caption{Pseudocode for embedding ERI transformation with FFTDF.} 
  \label{alg:eri with fftdf}
  \begin{algorithmic}[1]
    \For{all $\vecq$}
    \For{$\qty(\veck_{\mu}, \veck_{\nu})$ that conserves momentum} 
    \State Transform $\eri{\vecr}{\mu\veck_\mu \nu\veck_\nu}$ to $\eri{\vecr}{\tilde{i}\veck_\mu \tilde{j} \veck_\nu}$ by $C^{\veck, \AO, \EO}$
    \Comment{$\veck$-AO to $\veck$-EO}
    \State $\eri{\vecr}{\veczero \tilde{i} \veczero \tilde{j}} \pluseq \frac{1}{N_{\veck}} \eri{\vecr}{\tilde{i} \veck_\mu \tilde{j} \veck_\nu}$ \Comment{FT to the reference cell $\vecR = \veczero$}
    \EndFor
    \State Calculate $\eri{\vecG}{\veczero \tilde{i} \veczero \tilde{j}}$ using FFT
    \State $\eri{\vecG}{\veczero \tilde{i} \veczero \tilde{j}} \timeseq \frac{4\pi}{\Omega \qty|\vecq + \vecG|^2}$
    \State Calculate $\eri{\vecr}{\veczero \tilde{i} \veczero \tilde{j}}$ using inverse FFT
    \For{$\qty(\veck_{\kappa}, \veck_{\lambda})$ that conserves momentum} 
    \State Transform $\eri{\vecr}{\kappa\veck_\kappa \lambda\veck_\lambda}$ to $\eri{\vecr}{\tilde{k}\veck_\kappa \tilde{l} \veck_\lambda}$ by $C^{\veck, \AO, \EO}$
    \Comment{$\veck$-AO to $\veck$-EO}
    \State $\eri{\vecr}{\veczero \tilde{k} \veczero \tilde{l}} \pluseq \frac{1}{N_{\veck}} \eri{\vecr}{\tilde{k} \veck_\kappa \tilde{l} \veck_\lambda}$ \Comment{FT to the reference cell $\vecR = \veczero$}
    \EndFor  
    \State $\eri{\tilde{i}\tilde{j}}{\tilde{k}\tilde{l}} \pluseq \frac{1}{N_{\veck}} \sum_{\vecr} \eri{\veczero \tilde{i} \veczero \tilde{j}}{\vecr} \eri{\vecr}{\veczero \tilde{k} \veczero \tilde{l}}$
    \Comment{Contraction for the embedding ERI}
    \EndFor
  \end{algorithmic}
\end{algorithm}


\end{appendix}

\bibliography{refs-dmet-solid}

\providecommand{\latin}[1]{#1}
\makeatletter
\providecommand{\doi}
  {\begingroup\let\do\@makeother\dospecials
  \catcode`\{=1 \catcode`\}=2 \doi@aux}
\providecommand{\doi@aux}[1]{\endgroup\texttt{#1}}
\makeatother
\providecommand*\mcitethebibliography{\thebibliography}
\csname @ifundefined\endcsname{endmcitethebibliography}
  {\let\endmcitethebibliography\endthebibliography}{}
\begin{mcitethebibliography}{95}
\providecommand*\natexlab[1]{#1}
\providecommand*\mciteSetBstSublistMode[1]{}
\providecommand*\mciteSetBstMaxWidthForm[2]{}
\providecommand*\mciteBstWouldAddEndPuncttrue
  {\def\EndOfBibitem{\unskip.}}
\providecommand*\mciteBstWouldAddEndPunctfalse
  {\let\EndOfBibitem\relax}
\providecommand*\mciteSetBstMidEndSepPunct[3]{}
\providecommand*\mciteSetBstSublistLabelBeginEnd[3]{}
\providecommand*\EndOfBibitem{}
\mciteSetBstSublistMode{f}
\mciteSetBstMaxWidthForm{subitem}{(\alph{mcitesubitemcount})}
\mciteSetBstSublistLabelBeginEnd
  {\mcitemaxwidthsubitemform\space}
  {\relax}
  {\relax}

\bibitem[Imada \latin{et~al.}(1998)Imada, Fujimori, and Tokura]{Imada98}
Imada,~M.; Fujimori,~A.; Tokura,~Y. Metal-insulator Transitions. \emph{Rev.
  Mod. Phys.} \textbf{1998}, \emph{70}, 1039--1263\relax
\mciteBstWouldAddEndPuncttrue
\mciteSetBstMidEndSepPunct{\mcitedefaultmidpunct}
{\mcitedefaultendpunct}{\mcitedefaultseppunct}\relax
\EndOfBibitem
\bibitem[Dagotto(1994)]{Dagotto94}
Dagotto,~E. Correlated Electrons in High-Temperature Superconductors.
  \emph{Rev. Mod. Phys.} \textbf{1994}, \emph{66}, 763--840\relax
\mciteBstWouldAddEndPuncttrue
\mciteSetBstMidEndSepPunct{\mcitedefaultmidpunct}
{\mcitedefaultendpunct}{\mcitedefaultseppunct}\relax
\EndOfBibitem
\bibitem[Sachdev(2003)]{Sachdev03}
Sachdev,~S. Colloquium: Order and Quantum Phase Transitions in the Cuprate
  Superconductors. \emph{Rev. Mod. Phys.} \textbf{2003}, \emph{75}, 913\relax
\mciteBstWouldAddEndPuncttrue
\mciteSetBstMidEndSepPunct{\mcitedefaultmidpunct}
{\mcitedefaultendpunct}{\mcitedefaultseppunct}\relax
\EndOfBibitem
\bibitem[Lee \latin{et~al.}(2006)Lee, Nagaosa, and Wen]{Lee06RMPhightc}
Lee,~P.~A.; Nagaosa,~N.; Wen,~X.-G. {Doping a Mott Insulator: Physics of
  High-Temperature Superconductivity}. \emph{Rev. Mod. Phys.} \textbf{2006},
  \emph{78}, 17\relax
\mciteBstWouldAddEndPuncttrue
\mciteSetBstMidEndSepPunct{\mcitedefaultmidpunct}
{\mcitedefaultendpunct}{\mcitedefaultseppunct}\relax
\EndOfBibitem
\bibitem[Zgid and Chan(2011)Zgid, and Chan]{Zgid11}
Zgid,~D.; Chan,~G. K.-L. Dynamical Mean-field Theory from a Quantum Chemical
  Perspective. \emph{J. Chem. Phys.} \textbf{2011}, \emph{134}, 094115\relax
\mciteBstWouldAddEndPuncttrue
\mciteSetBstMidEndSepPunct{\mcitedefaultmidpunct}
{\mcitedefaultendpunct}{\mcitedefaultseppunct}\relax
\EndOfBibitem
\bibitem[Sun and Chan(2016)Sun, and Chan]{Sun16QET}
Sun,~Q.; Chan,~G. K.-L. {Quantum Embedding Theories}. \emph{Acc. Chem. Res.}
  \textbf{2016}, \emph{49}, 2705\relax
\mciteBstWouldAddEndPuncttrue
\mciteSetBstMidEndSepPunct{\mcitedefaultmidpunct}
{\mcitedefaultendpunct}{\mcitedefaultseppunct}\relax
\EndOfBibitem
\bibitem[Anderson(1961)]{Anderson61}
Anderson,~P.~W. Local Magnetic States in Metals. \emph{Phys. Rev.}
  \textbf{1961}, \emph{124}, 41 -- 53\relax
\mciteBstWouldAddEndPuncttrue
\mciteSetBstMidEndSepPunct{\mcitedefaultmidpunct}
{\mcitedefaultendpunct}{\mcitedefaultseppunct}\relax
\EndOfBibitem
\bibitem[Georges and Kotliar(1992)Georges, and Kotliar]{Georges92}
Georges,~A.; Kotliar,~G. Hubbard Model in Infinite Dimensions. \emph{Phys. Rev.
  B} \textbf{1992}, \emph{45}, 6479--6483\relax
\mciteBstWouldAddEndPuncttrue
\mciteSetBstMidEndSepPunct{\mcitedefaultmidpunct}
{\mcitedefaultendpunct}{\mcitedefaultseppunct}\relax
\EndOfBibitem
\bibitem[Georges \latin{et~al.}(1996)Georges, Kotliar, Krauth, and
  Rozenberg]{Georges96}
Georges,~A.; Kotliar,~G.; Krauth,~W.; Rozenberg,~M.~J. Dynamical Mean-field
  Theory of Strongly Correlated Fermion Systems and the Limit of Infinite
  Dimensions. \emph{Rev. Mod. Phys.} \textbf{1996}, \emph{68}, 13--125\relax
\mciteBstWouldAddEndPuncttrue
\mciteSetBstMidEndSepPunct{\mcitedefaultmidpunct}
{\mcitedefaultendpunct}{\mcitedefaultseppunct}\relax
\EndOfBibitem
\bibitem[Kotliar \latin{et~al.}(2006)Kotliar, Savrasov, Haule, Oudovenko,
  Parcollet, and Marianetti]{Kotliar06RMP}
Kotliar,~G.; Savrasov,~S.~Y.; Haule,~K.; Oudovenko,~V.~S.; Parcollet,~O.;
  Marianetti,~C.~A. {Electronic Structure Calculations with Dynamical
  Mean-field Theory}. \emph{{Rev. Mod. Phys.}} \textbf{2006}, \emph{78},
  865\relax
\mciteBstWouldAddEndPuncttrue
\mciteSetBstMidEndSepPunct{\mcitedefaultmidpunct}
{\mcitedefaultendpunct}{\mcitedefaultseppunct}\relax
\EndOfBibitem
\bibitem[Held(2007)]{Held2007}
Held,~K. {Electronic structure calculations using dynamical mean field theory}.
  \emph{Adv. Phys.} \textbf{2007}, \emph{56}, 829--926\relax
\mciteBstWouldAddEndPuncttrue
\mciteSetBstMidEndSepPunct{\mcitedefaultmidpunct}
{\mcitedefaultendpunct}{\mcitedefaultseppunct}\relax
\EndOfBibitem
\bibitem[Maier \latin{et~al.}(2005)Maier, Jarell, Pruschke, and
  Hettler]{Maier05RMP}
Maier,~T.; Jarell,~M.; Pruschke,~T.; Hettler,~M.~H. Quantum Cluster Theories.
  \emph{Rev. Mod. Phys.} \textbf{2005}, \emph{77}, 1027 -- 1080\relax
\mciteBstWouldAddEndPuncttrue
\mciteSetBstMidEndSepPunct{\mcitedefaultmidpunct}
{\mcitedefaultendpunct}{\mcitedefaultseppunct}\relax
\EndOfBibitem
\bibitem[Potthoff(2003)]{Potthoff03}
Potthoff,~M. Self-Energy-Functional Approach to Systems of Correlated
  Electrons. \emph{Eur. Phys. J. B} \textbf{2003}, \emph{32}, 429\relax
\mciteBstWouldAddEndPuncttrue
\mciteSetBstMidEndSepPunct{\mcitedefaultmidpunct}
{\mcitedefaultendpunct}{\mcitedefaultseppunct}\relax
\EndOfBibitem
\bibitem[S\'en\'echal(2008)]{Senechal08}
S\'en\'echal,~D. An Introduction to Quantum Cluster Methods. \emph{arXiv:
  0806.2690 [cond-mat]} \textbf{2008}, \relax
\mciteBstWouldAddEndPunctfalse
\mciteSetBstMidEndSepPunct{\mcitedefaultmidpunct}
{}{\mcitedefaultseppunct}\relax
\EndOfBibitem
\bibitem[Kananenka \latin{et~al.}(2015)Kananenka, Gull, and Zgid]{Kananenka15}
Kananenka,~A.~A.; Gull,~E.; Zgid,~D. Systematically Improvable Multiscale
  Solver for Correlated Electron Systems. \emph{Phys. Rev. B} \textbf{2015},
  \emph{91}, 121111\relax
\mciteBstWouldAddEndPuncttrue
\mciteSetBstMidEndSepPunct{\mcitedefaultmidpunct}
{\mcitedefaultendpunct}{\mcitedefaultseppunct}\relax
\EndOfBibitem
\bibitem[Rusakov \latin{et~al.}(2019)Rusakov, Iskakov, Tran, and
  Zgid]{Rusakov19}
Rusakov,~A.~A.; Iskakov,~S.; Tran,~L.~N.; Zgid,~D. Self-Energy Embedding Theory
  (SEET) for Periodic Systems. \emph{J. Chem. Theory Comput.} \textbf{2019},
  \emph{15}, 229\relax
\mciteBstWouldAddEndPuncttrue
\mciteSetBstMidEndSepPunct{\mcitedefaultmidpunct}
{\mcitedefaultendpunct}{\mcitedefaultseppunct}\relax
\EndOfBibitem
\bibitem[Biermann(2014)]{Biermann14JPCM}
Biermann,~S. Dynamical Screening Effects in Correlated Electron Materials: A
  Progress Report on Combined Many-Body Perturbation and Dynamical Mean Field
  Theory: $GW$+DMFT. \emph{J. Phys. : Condens. Matter} \textbf{2014},
  \emph{26}, 173202\relax
\mciteBstWouldAddEndPuncttrue
\mciteSetBstMidEndSepPunct{\mcitedefaultmidpunct}
{\mcitedefaultendpunct}{\mcitedefaultseppunct}\relax
\EndOfBibitem
\bibitem[Knizia and Chan(2012)Knizia, and Chan]{Knizia12}
Knizia,~G.; Chan,~G. K.-L. Density Matrix Embedding: A Simple Alternative to
  Dynamical Mean-Field Theory. \emph{Phys. Rev. Lett.} \textbf{2012},
  \emph{109}, 186404\relax
\mciteBstWouldAddEndPuncttrue
\mciteSetBstMidEndSepPunct{\mcitedefaultmidpunct}
{\mcitedefaultendpunct}{\mcitedefaultseppunct}\relax
\EndOfBibitem
\bibitem[Bulik \latin{et~al.}(2014)Bulik, Scuseria, and Dukelsky]{Bulik14}
Bulik,~I.~W.; Scuseria,~G.~E.; Dukelsky,~J. Density Matrix Embedding from
  Broken Symmetry Lattice Mean Fields. \emph{Phys. Rev. B} \textbf{2014},
  \emph{89}, 035140\relax
\mciteBstWouldAddEndPuncttrue
\mciteSetBstMidEndSepPunct{\mcitedefaultmidpunct}
{\mcitedefaultendpunct}{\mcitedefaultseppunct}\relax
\EndOfBibitem
\bibitem[Chen \latin{et~al.}(2014)Chen, Booth, Sharma, Knizia, and
  Chan]{Chen14dmethoneycomb}
Chen,~Q.; Booth,~G.~H.; Sharma,~S.; Knizia,~G.; Chan,~G. K.-L. Intermediate and
  Spin-Liquid Phase of the Half-Filled Honeycomb Hubbard Model. \emph{Phys.
  Rev. B} \textbf{2014}, \emph{89}, 165134\relax
\mciteBstWouldAddEndPuncttrue
\mciteSetBstMidEndSepPunct{\mcitedefaultmidpunct}
{\mcitedefaultendpunct}{\mcitedefaultseppunct}\relax
\EndOfBibitem
\bibitem[Zheng and Chan(2016)Zheng, and Chan]{Zheng16}
Zheng,~B.-X.; Chan,~G. K.-L. Ground-state Phase Diagram of the Square Lattice
  {Hubbard} Model from Density Matrix Embedding Theory. \emph{Phys. Rev. B}
  \textbf{2016}, \emph{93}, 035126\relax
\mciteBstWouldAddEndPuncttrue
\mciteSetBstMidEndSepPunct{\mcitedefaultmidpunct}
{\mcitedefaultendpunct}{\mcitedefaultseppunct}\relax
\EndOfBibitem
\bibitem[Zheng \latin{et~al.}(2017)Zheng, Kretchmer, Shi, Zhang, and
  Chan]{Zheng17}
Zheng,~B.-X.; Kretchmer,~J.~S.; Shi,~H.; Zhang,~S.; Chan,~G. K.-L. Cluster Size
  Convergence of the Density Matrix Embedding Theory and Its Dynamical Cluster
  Formulation: {A} Study with an Auxiliary-field Quantum {Monte} {Carlo}
  Solver. \emph{Phys. Rev. B} \textbf{2017}, \emph{95}, 045103\relax
\mciteBstWouldAddEndPuncttrue
\mciteSetBstMidEndSepPunct{\mcitedefaultmidpunct}
{\mcitedefaultendpunct}{\mcitedefaultseppunct}\relax
\EndOfBibitem
\bibitem[Wesolowski and Warshel(1993)Wesolowski, and Warshel]{Wesolowski93}
Wesolowski,~T.; Warshel,~A. {Frozen Density Functional Approach for Ab Initio
  Calculations of Solvated Molecules}. \emph{J. Phys. Chem} \textbf{1993},
  \emph{97}, 8050\relax
\mciteBstWouldAddEndPuncttrue
\mciteSetBstMidEndSepPunct{\mcitedefaultmidpunct}
{\mcitedefaultendpunct}{\mcitedefaultseppunct}\relax
\EndOfBibitem
\bibitem[Goodpaster \latin{et~al.}(2010)Goodpaster, Ananth, Manby, and
  Miller]{Goodpaster10dftemb}
Goodpaster,~J.~D.; Ananth,~N.; Manby,~F.~R.; Miller,~T.~F. Exact Nonadditive
  Kinetic Potentials for Embedded Density Functional Theory. \emph{J. Chem.
  Phys.} \textbf{2010}, \emph{133}, 084103\relax
\mciteBstWouldAddEndPuncttrue
\mciteSetBstMidEndSepPunct{\mcitedefaultmidpunct}
{\mcitedefaultendpunct}{\mcitedefaultseppunct}\relax
\EndOfBibitem
\bibitem[Huang \latin{et~al.}(2011)Huang, Pavone, and Carter]{Huang11dftemb}
Huang,~C.; Pavone,~M.; Carter,~E.~A. Quantum Mechanical Embedding Theory Based
  on a Unique Embedding Potential. \emph{J. Chem. Phys.} \textbf{2011},
  \emph{134}, 154110\relax
\mciteBstWouldAddEndPuncttrue
\mciteSetBstMidEndSepPunct{\mcitedefaultmidpunct}
{\mcitedefaultendpunct}{\mcitedefaultseppunct}\relax
\EndOfBibitem
\bibitem[Libisch \latin{et~al.}(2014)Libisch, Huang, and Carter]{Libisch14}
Libisch,~F.; Huang,~C.; Carter,~E.~A. {Embedded Correlated Wavefunction
  Schemes: Theory and Applications}. \emph{Acc. Chem. Res.} \textbf{2014},
  \emph{47}, 2768\relax
\mciteBstWouldAddEndPuncttrue
\mciteSetBstMidEndSepPunct{\mcitedefaultmidpunct}
{\mcitedefaultendpunct}{\mcitedefaultseppunct}\relax
\EndOfBibitem
\bibitem[Jacob and Neugebauer(2014)Jacob, and Neugebauer]{Jacob14}
Jacob,~C.~R.; Neugebauer,~J. {Subsystem Density‐Functional Theory}.
  \emph{WIREs Comput. Mol. Sci.} \textbf{2014}, \emph{4}, 325\relax
\mciteBstWouldAddEndPuncttrue
\mciteSetBstMidEndSepPunct{\mcitedefaultmidpunct}
{\mcitedefaultendpunct}{\mcitedefaultseppunct}\relax
\EndOfBibitem
\bibitem[Chulhai and Goodpaster(2018)Chulhai, and Goodpaster]{Chulhai2018}
Chulhai,~D.~V.; Goodpaster,~J.~D. Projection-based Correlated Wave Function in
  Density Functional Theory Embedding for Periodic Systems. \emph{J. Chem.
  Theory Comput.} \textbf{2018}, \emph{14}, 1928--1942\relax
\mciteBstWouldAddEndPuncttrue
\mciteSetBstMidEndSepPunct{\mcitedefaultmidpunct}
{\mcitedefaultendpunct}{\mcitedefaultseppunct}\relax
\EndOfBibitem
\bibitem[Lee \latin{et~al.}(2019)Lee, Welborn, Manby, and Miller]{Lee19dftemb}
Lee,~S. J.~R.; Welborn,~M.; Manby,~F.~R.; Miller,~T.~F. Projection-Based
  Wavefunction-in-DFT Embedding. \emph{Acc. Chem. Res.} \textbf{2019},
  \emph{52}, 1359\relax
\mciteBstWouldAddEndPuncttrue
\mciteSetBstMidEndSepPunct{\mcitedefaultmidpunct}
{\mcitedefaultendpunct}{\mcitedefaultseppunct}\relax
\EndOfBibitem
\bibitem[Zhu \latin{et~al.}(2016)Zhu, de~Silva, van Aggelen, and {Van
  Voorhis}]{Zhu2016}
Zhu,~T.; de~Silva,~P.; van Aggelen,~H.; {Van Voorhis},~T. {Many-Electron
  Expansion: A Density Functional Hierarchy for Strongly Correlated Systems}.
  \emph{Phys. Rev. B} \textbf{2016}, \emph{93}, 201108\relax
\mciteBstWouldAddEndPuncttrue
\mciteSetBstMidEndSepPunct{\mcitedefaultmidpunct}
{\mcitedefaultendpunct}{\mcitedefaultseppunct}\relax
\EndOfBibitem
\bibitem[Zhu \latin{et~al.}(2019)Zhu, de~Silva, and {Van Voorhis}]{Zhu2019a}
Zhu,~T.; de~Silva,~P.; {Van Voorhis},~T. {Implementation of the Many-Pair
  Expansion for Systematically Improving Density Functional Calculations of
  Molecules}. \emph{J. Chem. Theory Comput.} \textbf{2019}, \emph{15},
  1089--1101\relax
\mciteBstWouldAddEndPuncttrue
\mciteSetBstMidEndSepPunct{\mcitedefaultmidpunct}
{\mcitedefaultendpunct}{\mcitedefaultseppunct}\relax
\EndOfBibitem
\bibitem[Fan and Jie(2015)Fan, and Jie]{Fan15}
Fan,~Z.; Jie,~Q.-l. Cluster Density Matrix Embedding Theory for Quantum Spin
  Systems. \emph{Phys. Rev. B} \textbf{2015}, \emph{91}, 195118\relax
\mciteBstWouldAddEndPuncttrue
\mciteSetBstMidEndSepPunct{\mcitedefaultmidpunct}
{\mcitedefaultendpunct}{\mcitedefaultseppunct}\relax
\EndOfBibitem
\bibitem[Zheng \latin{et~al.}(2017)Zheng, Chung, Corboz, Ehlers, Qin, Noack,
  Shi, White, Zhang, and Chan]{Zheng17sci}
Zheng,~B.-X.; Chung,~C.-M.; Corboz,~P.; Ehlers,~G.; Qin,~M.-P.; Noack,~R.~M.;
  Shi,~H.; White,~S.~R.; Zhang,~S.; Chan,~G. K.-L. Stripe Order in the
  Underdoped Region of the Two-dimensional {Hubbard} Model. \emph{Science}
  \textbf{2017}, \emph{358}, 1155--1160\relax
\mciteBstWouldAddEndPuncttrue
\mciteSetBstMidEndSepPunct{\mcitedefaultmidpunct}
{\mcitedefaultendpunct}{\mcitedefaultseppunct}\relax
\EndOfBibitem
\bibitem[Gunst \latin{et~al.}(2017)Gunst, Wouters, De~Baerdemacker, and
  Van~Neck]{Gunst17}
Gunst,~K.; Wouters,~S.; De~Baerdemacker,~S.; Van~Neck,~D. Block Product Density
  Matrix Embedding Theory for Strongly Correlated Spin Systems. \emph{Phys.
  Rev. B} \textbf{2017}, \emph{95}, 195127\relax
\mciteBstWouldAddEndPuncttrue
\mciteSetBstMidEndSepPunct{\mcitedefaultmidpunct}
{\mcitedefaultendpunct}{\mcitedefaultseppunct}\relax
\EndOfBibitem
\bibitem[Sandhoefer and Chan(2016)Sandhoefer, and Chan]{Sandhoefer16}
Sandhoefer,~B.; Chan,~G. K.-L. Density Matrix Embedding Theory for Interacting
  Electron-phonon Systems. \emph{Phys. Rev. B} \textbf{2016}, \emph{94},
  085115\relax
\mciteBstWouldAddEndPuncttrue
\mciteSetBstMidEndSepPunct{\mcitedefaultmidpunct}
{\mcitedefaultendpunct}{\mcitedefaultseppunct}\relax
\EndOfBibitem
\bibitem[Wu \latin{et~al.}(2019)Wu, Cui, Tong, Lindsey, Chan, and
  Lin]{Wu19pdmet}
Wu,~X.; Cui,~Z.-H.; Tong,~Y.; Lindsey,~M.; Chan,~G. K.-L.; Lin,~L. Projected
  Density Matrix Embedding Theory with Applications to the Two-Dimensional
  Hubbard Model. \emph{J. Chem. Phys.} \textbf{2019}, \emph{151}, 064108\relax
\mciteBstWouldAddEndPuncttrue
\mciteSetBstMidEndSepPunct{\mcitedefaultmidpunct}
{\mcitedefaultendpunct}{\mcitedefaultseppunct}\relax
\EndOfBibitem
\bibitem[Knizia and Chan(2013)Knizia, and Chan]{Knizia13}
Knizia,~G.; Chan,~G. K.-L. Density {Matrix} {Embedding}: {A}
  {Strong}-{Coupling} {Quantum} {Embedding} {Theory}. \emph{J. Chem. Theory
  Comput.} \textbf{2013}, \emph{9}, 1428--1432\relax
\mciteBstWouldAddEndPuncttrue
\mciteSetBstMidEndSepPunct{\mcitedefaultmidpunct}
{\mcitedefaultendpunct}{\mcitedefaultseppunct}\relax
\EndOfBibitem
\bibitem[Wouters \latin{et~al.}(2016)Wouters, Jim{\'e}nez-Hoyos, Sun, and
  Chan]{Wouters16}
Wouters,~S.; Jim{\'e}nez-Hoyos,~C.~A.; Sun,~Q.; Chan,~G. K.-L. A {Practical}
  {Guide} to {Density} {Matrix} {Embedding} {Theory} in {Quantum} {Chemistry}.
  \emph{J. Chem. Theory Comput.} \textbf{2016}, \emph{12}, 2706--2719\relax
\mciteBstWouldAddEndPuncttrue
\mciteSetBstMidEndSepPunct{\mcitedefaultmidpunct}
{\mcitedefaultendpunct}{\mcitedefaultseppunct}\relax
\EndOfBibitem
\bibitem[Fulde and Stoll(2017)Fulde, and Stoll]{Fulde17dmet}
Fulde,~P.; Stoll,~H. {Dealing With the Exponential Wall in Electronic Structure
  Calculations}. \emph{J. Chem. Phys.} \textbf{2017}, \emph{146}, 194107\relax
\mciteBstWouldAddEndPuncttrue
\mciteSetBstMidEndSepPunct{\mcitedefaultmidpunct}
{\mcitedefaultendpunct}{\mcitedefaultseppunct}\relax
\EndOfBibitem
\bibitem[Pham \latin{et~al.}(2018)Pham, Bernales, and Gagliardi]{Pham18}
Pham,~H.~Q.; Bernales,~V.; Gagliardi,~L. Can {Density} {Matrix} {Embedding}
  {Theory} with the {Complete} {Activate} {Space} {Self}-{Consistent} {Field}
  {Solver} {Describe} {Single} and {Double} {Bond} {Breaking} in {Molecular}
  {Systems}? \emph{J. Chem. Theory Comput.} \textbf{2018}, \emph{14},
  1960--1968\relax
\mciteBstWouldAddEndPuncttrue
\mciteSetBstMidEndSepPunct{\mcitedefaultmidpunct}
{\mcitedefaultendpunct}{\mcitedefaultseppunct}\relax
\EndOfBibitem
\bibitem[Bulik \latin{et~al.}(2014)Bulik, Chen, and Scuseria]{Bulik14detsolid}
Bulik,~I.~W.; Chen,~W.; Scuseria,~G.~E. Electron Correlation in Solids via
  Density Embedding Theory. \emph{J. Chem. Phys.} \textbf{2014}, \emph{141},
  054113\relax
\mciteBstWouldAddEndPuncttrue
\mciteSetBstMidEndSepPunct{\mcitedefaultmidpunct}
{\mcitedefaultendpunct}{\mcitedefaultseppunct}\relax
\EndOfBibitem
\bibitem[Sun \latin{et~al.}(2018)Sun, Berkelbach, Blunt, Booth, Guo, Li, Liu,
  McClain, Sayfutyarova, Sharma, Wouters, and Chan]{Sun18pyscf}
Sun,~Q.; Berkelbach,~T.~C.; Blunt,~N.~S.; Booth,~G.~H.; Guo,~S.; Li,~Z.;
  Liu,~J.; McClain,~J.~D.; Sayfutyarova,~E.~R.; Sharma,~S.; Wouters,~S.;
  Chan,~G. K.-L. {PySCF}: the {Python}-based Simulations of Chemistry
  Framework. \emph{WIREs Comput. Mol. Sci.} \textbf{2018}, \emph{8},
  e1340\relax
\mciteBstWouldAddEndPuncttrue
\mciteSetBstMidEndSepPunct{\mcitedefaultmidpunct}
{\mcitedefaultendpunct}{\mcitedefaultseppunct}\relax
\EndOfBibitem
\bibitem[{McClain} \latin{et~al.}(2017){McClain}, Sun, Chan, and
  Berkelbach]{McClain17}
{McClain},~J.; Sun,~Q.; Chan,~G. K.-L.; Berkelbach,~T.~C. Gaussian-Based
  Coupled-Cluster Theory for the Ground-State and Band Structure of Solids.
  \emph{J. Chem. Theory Comput.} \textbf{2017}, \emph{13}, 1209--1218\relax
\mciteBstWouldAddEndPuncttrue
\mciteSetBstMidEndSepPunct{\mcitedefaultmidpunct}
{\mcitedefaultendpunct}{\mcitedefaultseppunct}\relax
\EndOfBibitem
\bibitem[Sun \latin{et~al.}(2017)Sun, Berkelbach, McClain, and Chan]{Sun17}
Sun,~Q.; Berkelbach,~T.~C.; McClain,~J.~D.; Chan,~G. K.-L. Gaussian and
  Plane-wave Mixed Density Fitting for Periodic Systems. \emph{J. Chem. Phys.}
  \textbf{2017}, \emph{147}, 164119\relax
\mciteBstWouldAddEndPuncttrue
\mciteSetBstMidEndSepPunct{\mcitedefaultmidpunct}
{\mcitedefaultendpunct}{\mcitedefaultseppunct}\relax
\EndOfBibitem
\bibitem[Pham \latin{et~al.}(2019)Pham, Hermes, and Gagliardi]{Pham19DMETsolid}
Pham,~H.~Q.; Hermes,~M.~R.; Gagliardi,~L. Periodic Electronic Structure
  Calculations With Density Matrix Embedding Theory. \emph{arXiv preprint
  arXiv:1909.08783} \textbf{2019}, \relax
\mciteBstWouldAddEndPunctfalse
\mciteSetBstMidEndSepPunct{\mcitedefaultmidpunct}
{}{\mcitedefaultseppunct}\relax
\EndOfBibitem
\bibitem[Foster and Boys(1960)Foster, and Boys]{Foster60}
Foster,~J.~M.; Boys,~S.~F. Canonical {Configurational} {Interaction}
  {Procedure}. \emph{Rev. Mod. Phys.} \textbf{1960}, \emph{32}, 300--302\relax
\mciteBstWouldAddEndPuncttrue
\mciteSetBstMidEndSepPunct{\mcitedefaultmidpunct}
{\mcitedefaultendpunct}{\mcitedefaultseppunct}\relax
\EndOfBibitem
\bibitem[Pipek and Mezey(1998)Pipek, and Mezey]{Pipek98}
Pipek,~J.; Mezey,~P.~G. A Fast Intrinsic Localization Procedure Applicable for
  Ab-initio and Semiempirical Linear Combination of Atomic Orbital Wave
  Functions. \emph{J. Chem. Phys.} \textbf{1998}, \emph{90}, 4916\relax
\mciteBstWouldAddEndPuncttrue
\mciteSetBstMidEndSepPunct{\mcitedefaultmidpunct}
{\mcitedefaultendpunct}{\mcitedefaultseppunct}\relax
\EndOfBibitem
\bibitem[Edmiston and Ruedenberg(1963)Edmiston, and Ruedenberg]{Edmiston63}
Edmiston,~C.; Ruedenberg,~K. Localized {Atomic} and {Molecular} {Orbitals}.
  \emph{Rev. Mod. Phys.} \textbf{1963}, \emph{35}, 457--464\relax
\mciteBstWouldAddEndPuncttrue
\mciteSetBstMidEndSepPunct{\mcitedefaultmidpunct}
{\mcitedefaultendpunct}{\mcitedefaultseppunct}\relax
\EndOfBibitem
\bibitem[Marzari and Vanderbilt(1997)Marzari, and Vanderbilt]{Marzari97}
Marzari,~N.; Vanderbilt,~D. {Maximally Localized Generalized Wannier Functions
  for Composite Energy Bands}. \emph{Phys. Rev. B} \textbf{1997}, \emph{56},
  12847--12865\relax
\mciteBstWouldAddEndPuncttrue
\mciteSetBstMidEndSepPunct{\mcitedefaultmidpunct}
{\mcitedefaultendpunct}{\mcitedefaultseppunct}\relax
\EndOfBibitem
\bibitem[Marzari \latin{et~al.}(2012)Marzari, Mostofi, Yates, Souza, and
  Vanderbilt]{Marzari12RMP}
Marzari,~N.; Mostofi,~A.~A.; Yates,~Y.~R.; Souza,~I.; Vanderbilt,~D. Maximally
  localized Wannier functions: Theory and applications. \emph{Rev. Mod. Phys.}
  \textbf{2012}, \emph{84}, 1419 -- 1475\relax
\mciteBstWouldAddEndPuncttrue
\mciteSetBstMidEndSepPunct{\mcitedefaultmidpunct}
{\mcitedefaultendpunct}{\mcitedefaultseppunct}\relax
\EndOfBibitem
\bibitem[J\'onsson \latin{et~al.}(2017)J\'onsson, Lehtola, Puska, and
  J\'onsson]{Jonsson17}
J\'onsson,~E.~{\"O}.; Lehtola,~S.; Puska,~M.; J\'onsson,~H. Theory and
  {Applications} of {Generalized} {Pipek}-{Mezey} {Wannier} {Functions}.
  \emph{J. Chem. Theory Comput.} \textbf{2017}, \emph{13}, 460--474\relax
\mciteBstWouldAddEndPuncttrue
\mciteSetBstMidEndSepPunct{\mcitedefaultmidpunct}
{\mcitedefaultendpunct}{\mcitedefaultseppunct}\relax
\EndOfBibitem
\bibitem[Souza \latin{et~al.}(2001)Souza, Marzari, and Vanderbilt]{Souza01}
Souza,~I.; Marzari,~N.; Vanderbilt,~D. Maximally Localized {Wannier} Functions
  for Entangled Energy Bands. \emph{Phys. Rev. B} \textbf{2001}, \emph{65},
  035109\relax
\mciteBstWouldAddEndPuncttrue
\mciteSetBstMidEndSepPunct{\mcitedefaultmidpunct}
{\mcitedefaultendpunct}{\mcitedefaultseppunct}\relax
\EndOfBibitem
\bibitem[Damle and Lin(2018)Damle, and Lin]{Damle18}
Damle,~A.; Lin,~L. Disentanglement via {Entanglement}: {A} {Unified} {Method}
  for {Wannier} {Localization}. \emph{Multiscale Model. Simul.} \textbf{2018},
  \emph{16}, 1392--1410\relax
\mciteBstWouldAddEndPuncttrue
\mciteSetBstMidEndSepPunct{\mcitedefaultmidpunct}
{\mcitedefaultendpunct}{\mcitedefaultseppunct}\relax
\EndOfBibitem
\bibitem[L\"owdin(1950)]{Lowdin50}
L\"owdin,~P.-O. On the {Non}-{Orthogonality} {Problem} {Connected} with the
  {Use} of {Atomic} {Wave} {Functions} in the {Theory} of {Molecules} and
  {Crystals}. \emph{J. Chem. Phys.} \textbf{1950}, \emph{18}, 365\relax
\mciteBstWouldAddEndPuncttrue
\mciteSetBstMidEndSepPunct{\mcitedefaultmidpunct}
{\mcitedefaultendpunct}{\mcitedefaultseppunct}\relax
\EndOfBibitem
\bibitem[Sun and Chan(2014)Sun, and Chan]{Sun14qmmm}
Sun,~Q.; Chan,~G. K.-L. Exact and Optimal Quantum Mechanics/Molecular Mechanics
  Boundaries. \emph{J. Chem. Theory Comput.} \textbf{2014}, \emph{10},
  3784\relax
\mciteBstWouldAddEndPuncttrue
\mciteSetBstMidEndSepPunct{\mcitedefaultmidpunct}
{\mcitedefaultendpunct}{\mcitedefaultseppunct}\relax
\EndOfBibitem
\bibitem[Reed \latin{et~al.}(1985)Reed, Weinstock, and Weinhold]{Reed85}
Reed,~A.~E.; Weinstock,~R.~B.; Weinhold,~F. Natural Population Analysis.
  \emph{J. Chem. Phys.} \textbf{1985}, \emph{83}, 735--746\relax
\mciteBstWouldAddEndPuncttrue
\mciteSetBstMidEndSepPunct{\mcitedefaultmidpunct}
{\mcitedefaultendpunct}{\mcitedefaultseppunct}\relax
\EndOfBibitem
\bibitem[Knizia(2013)]{Knizia13IAO}
Knizia,~G. Intrinsic {Atomic} {Orbitals}: {An} {Unbiased} {Bridge} between
  {Quantum} {Theory} and {Chemical} {Concepts}. \emph{J. Chem. Theory Comput.}
  \textbf{2013}, \emph{9}, 4834--4843\relax
\mciteBstWouldAddEndPuncttrue
\mciteSetBstMidEndSepPunct{\mcitedefaultmidpunct}
{\mcitedefaultendpunct}{\mcitedefaultseppunct}\relax
\EndOfBibitem
\bibitem[Saeb{\o} and Pulay(1993)Saeb{\o}, and Pulay]{Saebo93}
Saeb{\o},~S.; Pulay,~P. Local Treatment of Electron Correlation. \emph{Annu.
  Rev. Phys. Chem.} \textbf{1993}, \emph{44}, 213\relax
\mciteBstWouldAddEndPuncttrue
\mciteSetBstMidEndSepPunct{\mcitedefaultmidpunct}
{\mcitedefaultendpunct}{\mcitedefaultseppunct}\relax
\EndOfBibitem
\bibitem[Motta \latin{et~al.}(2017)Motta, Ceperley, Chan, Gomez, Gull, Guo,
  Jim{\'e}nez-Hoyos, Lan, Li, Ma, Millis, Prokof'ev, Ray, Scuseria, Sorella,
  Stoudenmire, Sun, Tupitsyn, White, Zgid, and Zhang]{Motta17}
Motta,~M.; Ceperley,~D.~M.; Chan,~G. K.-L.; Gomez,~J.~A.; Gull,~E.; Guo,~S.;
  Jim{\'e}nez-Hoyos,~C.~A.; Lan,~T.~N.; Li,~J.; Ma,~F.; Millis,~A.~J.;
  Prokof'ev,~N.~V.; Ray,~U.; Scuseria,~G.~E.; Sorella,~S.; Stoudenmire,~E.~M.;
  Sun,~Q.; Tupitsyn,~I.~S.; White,~S.~R.; Zgid,~D.; Zhang,~S. {Towards the
  Solution of the Many-Electron Problem in Real Materials: Equation of State of
  the Hydrogen Chain with State-of-the-Art Many-Body Methods}. \emph{Phys. Rev.
  X} \textbf{2017}, \emph{7}, 031059\relax
\mciteBstWouldAddEndPuncttrue
\mciteSetBstMidEndSepPunct{\mcitedefaultmidpunct}
{\mcitedefaultendpunct}{\mcitedefaultseppunct}\relax
\EndOfBibitem
\bibitem[Whitten(1973)]{Whitten73}
Whitten,~J.~L. Coulombic Potential Energy Integrals and Approximations.
  \emph{J. Chem. Phys.} \textbf{1973}, \emph{58}, 4496--4501\relax
\mciteBstWouldAddEndPuncttrue
\mciteSetBstMidEndSepPunct{\mcitedefaultmidpunct}
{\mcitedefaultendpunct}{\mcitedefaultseppunct}\relax
\EndOfBibitem
\bibitem[Savrasov \latin{et~al.}(2001)Savrasov, Kotliar, and
  Abrahams]{Savrasov01}
Savrasov,~S.~Y.; Kotliar,~G.; Abrahams,~E. {Correlated Electrons in
  $\delta$-Plutonium within a Dynamical Mean-Field Picture}. \emph{Nature}
  \textbf{2001}, \emph{410}, 793\relax
\mciteBstWouldAddEndPuncttrue
\mciteSetBstMidEndSepPunct{\mcitedefaultmidpunct}
{\mcitedefaultendpunct}{\mcitedefaultseppunct}\relax
\EndOfBibitem
\bibitem[Savrasov and Kotliar(2004)Savrasov, and Kotliar]{Savrasov04}
Savrasov,~S.~Y.; Kotliar,~G. {Spectral Density Functionals for Electronic
  Structure Calculations}. \emph{Phys. Rev. B} \textbf{2004}, \emph{65},
  245101\relax
\mciteBstWouldAddEndPuncttrue
\mciteSetBstMidEndSepPunct{\mcitedefaultmidpunct}
{\mcitedefaultendpunct}{\mcitedefaultseppunct}\relax
\EndOfBibitem
\bibitem[Pourovskii \latin{et~al.}(2007)Pourovskii, Amadon, Biermann, and
  Georges]{Pourovskii07}
Pourovskii,~L.~V.; Amadon,~B.; Biermann,~S.; Georges,~A. {Self-Consistency Over
  the Charge Density in Dynamical Mean-Field Theory: A Linear Muffin-tin
  Implementation and Some Physical Implications}. \emph{{Phys. Rev. B}}
  \textbf{2007}, \emph{76}, 235101\relax
\mciteBstWouldAddEndPuncttrue
\mciteSetBstMidEndSepPunct{\mcitedefaultmidpunct}
{\mcitedefaultendpunct}{\mcitedefaultseppunct}\relax
\EndOfBibitem
\bibitem[Park \latin{et~al.}(2014)Park, Millis, and Marianetti]{Park14dmftcsc}
Park,~W.; Millis,~A.~J.; Marianetti,~C.~A. Computing Total Energies in Complex
  Materials Using Charge Self-Consistent DFT + DMFT. \emph{Phys. Rev. B}
  \textbf{2014}, \emph{90}, 235103\relax
\mciteBstWouldAddEndPuncttrue
\mciteSetBstMidEndSepPunct{\mcitedefaultmidpunct}
{\mcitedefaultendpunct}{\mcitedefaultseppunct}\relax
\EndOfBibitem
\bibitem[Li \latin{et~al.}(2011)Li, Chen, Behan, Zhang, Petravic, and
  Glushenkov]{Li11hBN}
Li,~L.~H.; Chen,~Y.; Behan,~G.; Zhang,~H.; Petravic,~M.; Glushenkov,~A.~M.
  Large-Scale Mechanical Peeling of Boron Nitride Nanosheets by Low-energy Ball
  Milling. \emph{J. Mater. Chem.} \textbf{2011}, \emph{21}, 11862\relax
\mciteBstWouldAddEndPuncttrue
\mciteSetBstMidEndSepPunct{\mcitedefaultmidpunct}
{\mcitedefaultendpunct}{\mcitedefaultseppunct}\relax
\EndOfBibitem
\bibitem[T{\"{o}}bbens \latin{et~al.}(2001)T{\"{o}}bbens, St{\"{u}}{\ss }er,
  Knorr, Mayer, and Lampert]{Tobbens01Si}
T{\"{o}}bbens,~D.; St{\"{u}}{\ss }er,~N.; Knorr,~K.; Mayer,~H.; Lampert,~G. E9:
  The New High-Resolution Neutron Powder Diffractometer at the Berlin Neutron
  Scattering Center. European Powder Diffraction EPDIC 7. 2001; pp
  288--293\relax
\mciteBstWouldAddEndPuncttrue
\mciteSetBstMidEndSepPunct{\mcitedefaultmidpunct}
{\mcitedefaultendpunct}{\mcitedefaultseppunct}\relax
\EndOfBibitem
\bibitem[Cheetham and Hope(1983)Cheetham, and Hope]{Cheetham83NiO}
Cheetham,~A.~K.; Hope,~D. A.~O. Magnetic Ordering and Exchange Effects in the
  Antiferromagnetic Solid Solutions \ce{Mn_{x}Ni_{1-x}O}. \emph{Phys. Rev. B}
  \textbf{1983}, \emph{27}, 6964\relax
\mciteBstWouldAddEndPuncttrue
\mciteSetBstMidEndSepPunct{\mcitedefaultmidpunct}
{\mcitedefaultendpunct}{\mcitedefaultseppunct}\relax
\EndOfBibitem
\bibitem[Perdew \latin{et~al.}(1996)Perdew, Burke, and Ernzerhof]{Perdew96PBE}
Perdew,~J.~P.; Burke,~K.; Ernzerhof,~M. {Generalized Gradient Approximation
  Made Simple}. \emph{{Phys. Rev. Lett.}} \textbf{1996}, \emph{77}, 3865\relax
\mciteBstWouldAddEndPuncttrue
\mciteSetBstMidEndSepPunct{\mcitedefaultmidpunct}
{\mcitedefaultendpunct}{\mcitedefaultseppunct}\relax
\EndOfBibitem
\bibitem[Goedecker \latin{et~al.}(1996)Goedecker, Teter, and
  Hutter]{Goedecker96}
Goedecker,~S.; Teter,~M.; Hutter,~J. Separable Dual-Space Gaussian
  Pseudopotentials. \emph{Phys. Rev. B} \textbf{1996}, \emph{54}, 1703\relax
\mciteBstWouldAddEndPuncttrue
\mciteSetBstMidEndSepPunct{\mcitedefaultmidpunct}
{\mcitedefaultendpunct}{\mcitedefaultseppunct}\relax
\EndOfBibitem
\bibitem[Hartwigsen \latin{et~al.}(1998)Hartwigsen, Goedecker, and
  Hutter]{Hartwigsen98}
Hartwigsen,~C.; Goedecker,~S.; Hutter,~J. Relativistic Separable Dual-Space
  Gaussian Pseudopotentials from H to Rn. \emph{Phys. Rev. B} \textbf{1998},
  \emph{58}, 3641\relax
\mciteBstWouldAddEndPuncttrue
\mciteSetBstMidEndSepPunct{\mcitedefaultmidpunct}
{\mcitedefaultendpunct}{\mcitedefaultseppunct}\relax
\EndOfBibitem
\bibitem[VandeVondele and Hutter(2007)VandeVondele, and
  Hutter]{VandeVondele2007}
VandeVondele,~J.; Hutter,~J. {Gaussian Basis Sets for Accurate Calculations on
  Molecular Systems in Gas and Condensed Phases}. \emph{J. Chem. Phys.}
  \textbf{2007}, \emph{127}, 114105\relax
\mciteBstWouldAddEndPuncttrue
\mciteSetBstMidEndSepPunct{\mcitedefaultmidpunct}
{\mcitedefaultendpunct}{\mcitedefaultseppunct}\relax
\EndOfBibitem
\bibitem[Stoychev \latin{et~al.}(2017)Stoychev, Auer, and
  Neese]{Stoychev17ETBbasis}
Stoychev,~G.~L.; Auer,~A.~A.; Neese,~F. Automatic Generation of Auxiliary Basis
  Sets. \emph{J. Chem. Theory Comput.} \textbf{2017}, \emph{13}, 554--562\relax
\mciteBstWouldAddEndPuncttrue
\mciteSetBstMidEndSepPunct{\mcitedefaultmidpunct}
{\mcitedefaultendpunct}{\mcitedefaultseppunct}\relax
\EndOfBibitem
\bibitem[Paier \latin{et~al.}(2005)Paier, Hirschl, Marsman, and
  Kresse]{Paier05}
Paier,~J.; Hirschl,~R.; Marsman,~M.; Kresse,~G. The Perdew-Burke-Ernzerhof
  Exchange-Correlation Functional Applied to the G2-1 Test Set Using a
  Plane-wave Basis Set. \emph{J. Chem. Phys.} \textbf{2005}, \emph{122},
  234102\relax
\mciteBstWouldAddEndPuncttrue
\mciteSetBstMidEndSepPunct{\mcitedefaultmidpunct}
{\mcitedefaultendpunct}{\mcitedefaultseppunct}\relax
\EndOfBibitem
\bibitem[Sundararaman and Arias(2013)Sundararaman, and
  Arias]{Sundararaman13regularization}
Sundararaman,~R.; Arias,~T. Regularization of the Coulomb Singularity in Exact
  Exchange by Wigner-Seitz Truncated Interactions: Towards Chemical Accuracy in
  Nontrivial Systems. \emph{Phys. Rev. B} \textbf{2013}, \emph{87},
  165122\relax
\mciteBstWouldAddEndPuncttrue
\mciteSetBstMidEndSepPunct{\mcitedefaultmidpunct}
{\mcitedefaultendpunct}{\mcitedefaultseppunct}\relax
\EndOfBibitem
\bibitem[Bartlett and Musia{\l}(2007)Bartlett, and Musia{\l}]{Bartlett07}
Bartlett,~J.~R.; Musia{\l},~M. Coupled-Cluster Theory in Quantum Chemistry.
  \emph{Rev. Mod. Phys.} \textbf{2007}, \emph{79}, 291\relax
\mciteBstWouldAddEndPuncttrue
\mciteSetBstMidEndSepPunct{\mcitedefaultmidpunct}
{\mcitedefaultendpunct}{\mcitedefaultseppunct}\relax
\EndOfBibitem
\bibitem[Shavitt and Bartlett(2009)Shavitt, and Bartlett]{Shavitt09book}
Shavitt,~I.; Bartlett,~R.~J. \emph{Many-Body Methods in Chemistry and Physics:
  MBPT and Coupled Cluster Theory}; Cambridge University, 2009\relax
\mciteBstWouldAddEndPuncttrue
\mciteSetBstMidEndSepPunct{\mcitedefaultmidpunct}
{\mcitedefaultendpunct}{\mcitedefaultseppunct}\relax
\EndOfBibitem
\bibitem[Gao \latin{et~al.}(2019)Gao, Sun, Yu, Motta, McClain, White, Minnich,
  and Chan]{Gao19CCTMO}
Gao,~Y.; Sun,~Q.; Yu,~J.~M.; Motta,~M.; McClain,~J.; White,~A.~F.;
  Minnich,~A.~J.; Chan,~G.~K. Electronic Structure of Bulk Manganese Oxide and
  Nickel Oxide from Coupled Cluster Theory. \emph{arXiv preprint
  arXiv:1910.02191} \textbf{2019}, \relax
\mciteBstWouldAddEndPunctfalse
\mciteSetBstMidEndSepPunct{\mcitedefaultmidpunct}
{}{\mcitedefaultseppunct}\relax
\EndOfBibitem
\bibitem[Zhang and Gr{\"u}neis(2019)Zhang, and Gr{\"u}neis]{Zhang19KCCreview}
Zhang,~I.~Y.; Gr{\"u}neis,~A. Coupled Cluster Theory in Materials Science.
  \emph{Front. Mater.} \textbf{2019}, \emph{6}, 123\relax
\mciteBstWouldAddEndPuncttrue
\mciteSetBstMidEndSepPunct{\mcitedefaultmidpunct}
{\mcitedefaultendpunct}{\mcitedefaultseppunct}\relax
\EndOfBibitem
\bibitem[Gygi and Baldereschi(1986)Gygi, and Baldereschi]{Gygi86}
Gygi,~F.; Baldereschi,~A. {Self-Consistent Hartree-Fock and Screened-Exchange
  Calculations in Solids: Applications to Silicon}. \emph{{Phys. Rev. B}}
  \textbf{1986}, \emph{34}, 4405\relax
\mciteBstWouldAddEndPuncttrue
\mciteSetBstMidEndSepPunct{\mcitedefaultmidpunct}
{\mcitedefaultendpunct}{\mcitedefaultseppunct}\relax
\EndOfBibitem
\bibitem[Welborn \latin{et~al.}(2016)Welborn, Tsuchimochi, and
  Van~Voorhis]{Welborn16bootstrap}
Welborn,~M.; Tsuchimochi,~T.; Van~Voorhis,~T. {Bootstrap Embedding: An
  Internally Consistent Fragment-Based Method}. \emph{J. Chem. Phys.}
  \textbf{2016}, \emph{145}, 074102\relax
\mciteBstWouldAddEndPuncttrue
\mciteSetBstMidEndSepPunct{\mcitedefaultmidpunct}
{\mcitedefaultendpunct}{\mcitedefaultseppunct}\relax
\EndOfBibitem
\bibitem[Ricke \latin{et~al.}(2017)Ricke, Welborn, Ye, and
  Van~Voorhis]{Ricke17bootstrap}
Ricke,~N.; Welborn,~M.; Ye,~H.-Z.; Van~Voorhis,~T. {Performance of Bootstrap
  Embedding for Long-Range Interactions and 2D Systems}. \emph{Mol. Phys.}
  \textbf{2017}, \emph{115}, 2242\relax
\mciteBstWouldAddEndPuncttrue
\mciteSetBstMidEndSepPunct{\mcitedefaultmidpunct}
{\mcitedefaultendpunct}{\mcitedefaultseppunct}\relax
\EndOfBibitem
\bibitem[Ye \latin{et~al.}(2019)Ye, Ricke, Tran, and
  Van~Voorhis]{Ye19bootstrap}
Ye,~H.-Z.; Ricke,~N.~D.; Tran,~H.~K.; Van~Voorhis,~T. Bootstrap Embedding for
  Molecules. \emph{J. Chem. Theory Comput.} \textbf{2019}, \emph{15},
  4497\relax
\mciteBstWouldAddEndPuncttrue
\mciteSetBstMidEndSepPunct{\mcitedefaultmidpunct}
{\mcitedefaultendpunct}{\mcitedefaultseppunct}\relax
\EndOfBibitem
\bibitem[Murnaghan(1944)]{Murnaghan44}
Murnaghan,~F.~D. The {Compressibility} of {Media} under {Extreme} {Pressures}.
  \emph{Proc. Natl. Acad. Sci. U S A} \textbf{1944}, \emph{30}, 244\relax
\mciteBstWouldAddEndPuncttrue
\mciteSetBstMidEndSepPunct{\mcitedefaultmidpunct}
{\mcitedefaultendpunct}{\mcitedefaultseppunct}\relax
\EndOfBibitem
\bibitem[Birch(1947)]{Birch47}
Birch,~F. Finite Elastic Strain of Cubic Crystals. \emph{Phys. Rev.}
  \textbf{1947}, \emph{71}, 809\relax
\mciteBstWouldAddEndPuncttrue
\mciteSetBstMidEndSepPunct{\mcitedefaultmidpunct}
{\mcitedefaultendpunct}{\mcitedefaultseppunct}\relax
\EndOfBibitem
\bibitem[Schimka \latin{et~al.}(2011)Schimka, Harl, and Kresse]{Schimka11}
Schimka,~L.; Harl,~J.; Kresse,~G. Improved Hybrid Functional for Solids: The
  HSEsol Functional. \emph{J. Chem. Phys.} \textbf{2011}, \emph{134},
  024116\relax
\mciteBstWouldAddEndPuncttrue
\mciteSetBstMidEndSepPunct{\mcitedefaultmidpunct}
{\mcitedefaultendpunct}{\mcitedefaultseppunct}\relax
\EndOfBibitem
\bibitem[Sawatzky and Allen(1984)Sawatzky, and Allen]{Sawatzky84}
Sawatzky,~G.~A.; Allen,~J.~W. {Magnitude and Origin of the Band Gap in NiO}.
  \emph{{Phys. Rev. Lett.}} \textbf{1984}, \emph{53}, 2339\relax
\mciteBstWouldAddEndPuncttrue
\mciteSetBstMidEndSepPunct{\mcitedefaultmidpunct}
{\mcitedefaultendpunct}{\mcitedefaultseppunct}\relax
\EndOfBibitem
\bibitem[Perdew \latin{et~al.}(2017)Perdew, Yang, Burke, Yang, Gross,
  Scheffler, Scuseria, Henderson, Zhang, Ruzsinszky, Peng, Sun, Trushin, and
  G{\"o}rling]{Perdew17}
Perdew,~J.~P.; Yang,~W.; Burke,~K.; Yang,~Z.; Gross,~E. K.~U.; Scheffler,~M.;
  Scuseria,~G.~E.; Henderson,~T.~M.; Zhang,~I.~Y.; Ruzsinszky,~A.; Peng,~H.;
  Sun,~J.; Trushin,~E.; G{\"o}rling,~A. Understanding {Band} {Gaps} of {Solids}
  in {Generalized} {Kohn}-{Sham} {Theory}. \emph{Proc. Natl. Acad. Sci. USA}
  \textbf{2017}, \emph{114}, 2801--2806\relax
\mciteBstWouldAddEndPuncttrue
\mciteSetBstMidEndSepPunct{\mcitedefaultmidpunct}
{\mcitedefaultendpunct}{\mcitedefaultseppunct}\relax
\EndOfBibitem
\bibitem[K\"ummel and Kronik(2008)K\"ummel, and Kronik]{Kuemmel08RMP}
K\"ummel,~S.; Kronik,~L. {Orbital-Dependent Density Functionals: Theory and
  Applications}. \emph{Rev. Mod. Phys.} \textbf{2008}, \emph{80}, 3\relax
\mciteBstWouldAddEndPuncttrue
\mciteSetBstMidEndSepPunct{\mcitedefaultmidpunct}
{\mcitedefaultendpunct}{\mcitedefaultseppunct}\relax
\EndOfBibitem
\bibitem[Alperin(1962)]{Alperin62NiOm}
Alperin,~A.~H. \emph{J. Phys. Soc. Japan Suppl. B} \textbf{1962}, \emph{17},
  12\relax
\mciteBstWouldAddEndPuncttrue
\mciteSetBstMidEndSepPunct{\mcitedefaultmidpunct}
{\mcitedefaultendpunct}{\mcitedefaultseppunct}\relax
\EndOfBibitem
\bibitem[Fender \latin{et~al.}(1968)Fender, Jacobson, and
  Wedgwood]{Fender68NiOm}
Fender,~B. E.~F.; Jacobson,~A.~J.; Wedgwood,~F.~A. Covalency Parameters in MnO,
  $\alpha$‐MnS, and NiO. \emph{J. Chem. Phys.} \textbf{1968}, \emph{48},
  990\relax
\mciteBstWouldAddEndPuncttrue
\mciteSetBstMidEndSepPunct{\mcitedefaultmidpunct}
{\mcitedefaultendpunct}{\mcitedefaultseppunct}\relax
\EndOfBibitem
\bibitem[Chatterji \latin{et~al.}(2009)Chatterji, McIntyre, and
  Lindgard]{Chatterji09}
Chatterji,~T.; McIntyre,~G.~J.; Lindgard,~P.-A. {Antiferromagnetic Phase
  Transition and Spin Correlations in NiO}. \emph{Phys. Rev. B} \textbf{2009},
  \emph{79}, 172403\relax
\mciteBstWouldAddEndPuncttrue
\mciteSetBstMidEndSepPunct{\mcitedefaultmidpunct}
{\mcitedefaultendpunct}{\mcitedefaultseppunct}\relax
\EndOfBibitem
\bibitem[Germann \latin{et~al.}(1974)Germann, Maier, and Strau{\ss}]{Germann74}
Germann,~K.~H.; Maier,~K.; Strau{\ss},~E. {Magnetic Order Induced Birefringence
  and Critical Behaviour of the Long Range Order Parameter in NiO}. \emph{Solid
  State Commun.} \textbf{1974}, \emph{14}, 1309\relax
\mciteBstWouldAddEndPuncttrue
\mciteSetBstMidEndSepPunct{\mcitedefaultmidpunct}
{\mcitedefaultendpunct}{\mcitedefaultseppunct}\relax
\EndOfBibitem
\bibitem[Negoveti{\'c} and Konstantinovi{\'c}(1973)Negoveti{\'c}, and
  Konstantinovi{\'c}]{Negovetic73}
Negoveti{\'c},~I.; Konstantinovi{\'c},~J. {The Critical Behaviour of
  Spontaneous Magnetization in the Antiferromagnetic NiO}. \emph{Solid State
  Commun.} \textbf{1973}, \emph{13}, 249\relax
\mciteBstWouldAddEndPuncttrue
\mciteSetBstMidEndSepPunct{\mcitedefaultmidpunct}
{\mcitedefaultendpunct}{\mcitedefaultseppunct}\relax
\EndOfBibitem
\bibitem[Zhu \latin{et~al.}(2019)Zhu, Cui, and Chan]{Zhu19-dmft-solid}
Zhu,~T.; Cui,~Z.-H.; Chan,~G. K.-L. Efficient Implementation of Ab Initio
  Quantum Embedding in Periodic Systems: Dynamical Mean-Field Theory.
  \emph{arXiv:1909.08592} \textbf{2019}, \relax
\mciteBstWouldAddEndPunctfalse
\mciteSetBstMidEndSepPunct{\mcitedefaultmidpunct}
{}{\mcitedefaultseppunct}\relax
\EndOfBibitem
\end{mcitethebibliography}

\end{document}